# Rachel: A general-purpose language model directs and revises retrosynthetic routes

*Qisheng Li*[1,2]*, Shunchao Jiang*[3]*, Chen Qi*[1]*, Xin Su*[4,*]*, Da Han*[5,6,*]*, Guangyong Chen*[1,*]

[1] Hangzhou Institute of Medicine, Chinese Academy of Sciences, Hangzhou, China

[2] Beijing Advanced Innovation Center for Soft Matter Science and Engineering and State Key Laboratory of Organic–Inorganic Composites, Beijing Key Laboratory of Bioprocess, College of Life Science and Technology, Beijing University of Chemical Technology, Beijing, China

[3] Zhenzhida Biotechnology (Shanghai) Co., Ltd.

[4] State Key Laboratory of Organic–Inorganic Composites, Beijing Advanced Innovation Center for Soft Matter Science and Engineering, College of Life Science and Technology, Beijing University of Chemical Technology, Beijing 100029, China

[5] Zhejiang Key Laboratory of Functional Nucleic Acids for Basic and Clinical Application, Hangzhou Institute of Medicine, Chinese Academy of Sciences, Hangzhou, China

[6] Institute of Molecular Medicine, Renji Hospital, School of Medicine, Shanghai Jiao Tong University, Shanghai, China

* Correspondence: Xin Su: xinsu@mail.buct.edu.cn; Da Han: dahan@sjtu.edu.cn; Guangyong Chen: gychen@link.cuhk.edu.hk.

**Abstract**

Retrosynthetic planning advances through decisions that reshape the remaining chemical problem: a locally plausible disconnection can leave precursors whose chemoselectivity constraints complicate the rest of the route. Existing planners often channel model proposals through search or template procedures, leaving open whether a general-purpose large language model (LLM) can itself sustain and revise route strategy. We developed Rachel, a stateful environment that executes and checks LLM-directed chemistry but prescribes neither a search policy nor a stopping rule. Without supplied reference routes or route-level solutions, GPT-5.5 achieved strict closure for 111 of 120 PaRoutes120 targets and 24 of 25 targets in the separate RF25 difficult-target cohort. RF25 was drawn largely from studies published after GPT-5.5's reported knowledge cutoff. Closure required complete routes and independent source resolution of every terminal precursor after planning. On a shared PaRoutes subset, forward-model support exceeded that of most comparator methods, and Rachel received the highest mean overall route score from both method-blinded LLM evaluators. Recorded trajectories showed continued model-proposed chemistry, with revised strategies carried into subsequent steps. Replacing LLM route decisions with fixed policies reduced strict closure to 6–15/120 despite continued local chemical execution; restricting planning support also reduced closure in RF25. Within Rachel, a general-purpose LLM coordinated successive chemical choices and revised its strategy as earlier decisions reshaped the remaining problems.

## Introduction

Retrosynthetic planning shapes how readily a designed structure can be obtained for study, whether the target is a drug candidate, a chemical-biology probe or a functional material.[1] Each accepted transformation defines the precursor structures that subsequent planning must address, and their functional groups and chemoselectivity constraints shape the available synthetic options. A disconnection that is chemically reasonable in isolation can therefore leave precursors that are difficult to trace back to suitable starting materials.[2,3] A planner must reassess the transformations still to come, and maintain or revise the overall strategy, as earlier decisions reshape the molecular problems that remain.

Contemporary computer-aided synthesis-planning systems commonly combine learned precursor generation with algorithmic route search. Reaction models propose candidate transformations, while search procedures determine which branches to expand and when route construction should stop.[4,5,6] The resulting route therefore depends jointly on local reaction models, search policies, stock definitions and stopping criteria.[6,7,8] Recent methods have incorporated route context into prediction or generated multistep routes directly as structured outputs.[9,10,11] Yet stronger single-step prediction does not necessarily translate into better route-finding performance,[8] indicating that multistep planning also depends on how chemical decisions are selected, carried forward and revised as a route develops.

Large language models have begun to open new perspectives on how synthesis planning might be organized.[12,13,14] Recent studies have explored LLMs in several multistep settings: DeepRetro combines model-proposed disconnections with chemical checks and recursive planning,[15] AOT* maps LLM-generated pathways into systematic AND–OR tree search,[16] and SyntheLite uses an LLM-generated synthesis blueprint to guide template-based route search.[17] These approaches couple LLM proposals or synthesis blueprints to recursive planning, systematic search or template-based refinement. A central question is whether a general-purpose LLM can direct successive route decisions and revise its strategy as accepted transformations change the molecular problems that remain. Answering this question requires tracing how the model responds to the chemical consequences of its decisions, together with controlled interventions that test their contribution to route construction.

In AI-assisted formal theorem proving, an explicit problem state lets a model propose a step, have it checked, and continue from the result, turning long reasoning into inspectable and cumulative transitions.[18,19] Chemistry admits no equivalent deductive guarantee: structural and reaction-level checks constrain a proposed transformation without establishing experimental feasibility.[20] A similar cycle can connect model decisions to executable chemical transformations, with accepted outcomes defining the next molecular problem.

We therefore developed Rachel, an executable and stateful chemical-planning environment for studying how a general-purpose LLM directs and revises a synthesis route (Fig. 1). Starting from a target molecule, the LLM proposes and compares transformations, decides which molecules to develop further or accept as terminals, and maintains or revises its strategy as the route changes. Rachel supplies molecular context, executable operations and validation feedback, and records accepted transformations and terminal decisions as persistent route states (Fig. 2). After planning ends, routes are frozen and their terminal structures undergo an independent source-availability audit. Strict closure requires a complete route and source resolution of every terminal precursor. With GPT-5.5, Rachel reached strict closure for 111 of 120 PaRoutes120 targets and 24 of 25 targets in the separate RF25 difficult-target cohort. Complementary computational assessments examined the proposed chemistry, while trajectory analyses traced how route strategies evolved. Replacing the LLM's route decisions with fixed policies using Rachel's candidate execution and validation procedures reduced strict closure to 6–15 of 120 targets; restricting planning support on RF25 reduced it to 2–18 of 25. Together, these records and interventions allow route construction to be examined through successive chemical choices, their consequences and subsequent revisions.

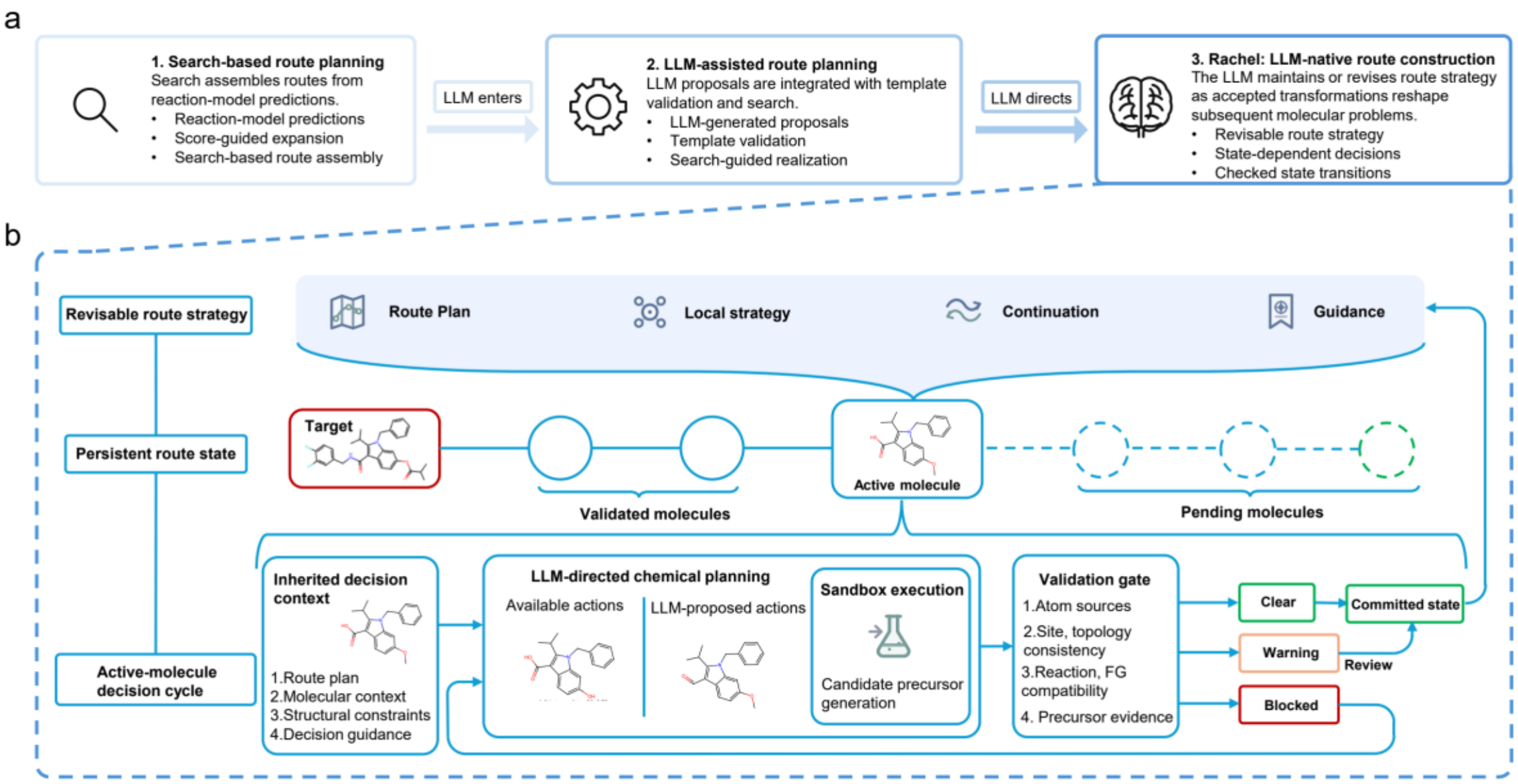

**Fig. 1 | From search-based route planning to LLM-native route construction in Rachel. a,** Three approaches to organizing retrosynthetic routes. Search-based planning assembles routes from reaction-model predictions. The illustrated LLM-assisted approaches combine model proposals with template validation and search. In Rachel, the LLM directs successive chemical decisions, maintaining or revising its strategy as accepted transformations reshape the molecular problems that remain. This comparison highlights differences in route organization, not performance.

**b,** Rachel's stateful chemical-planning workflow. A revisable route strategy guides decisions for each active molecule. Using the inherited molecular context and structural constraints, the LLM selects an available action or proposes a new transformation. Rachel executes the candidate action in a sandbox and checks the resulting precursors for atom sources, site and topology consistency, reaction and functional-group compatibility, and supporting evidence. Candidates with warnings return to the LLM for review; blocked candidates require reconsideration. Accepted transformations update the stored route state, and the resulting molecular context informs the next decision. The validation gate operates during planning. A separate terminal-source audit follows route freezing and is not shown. FG, functional group.

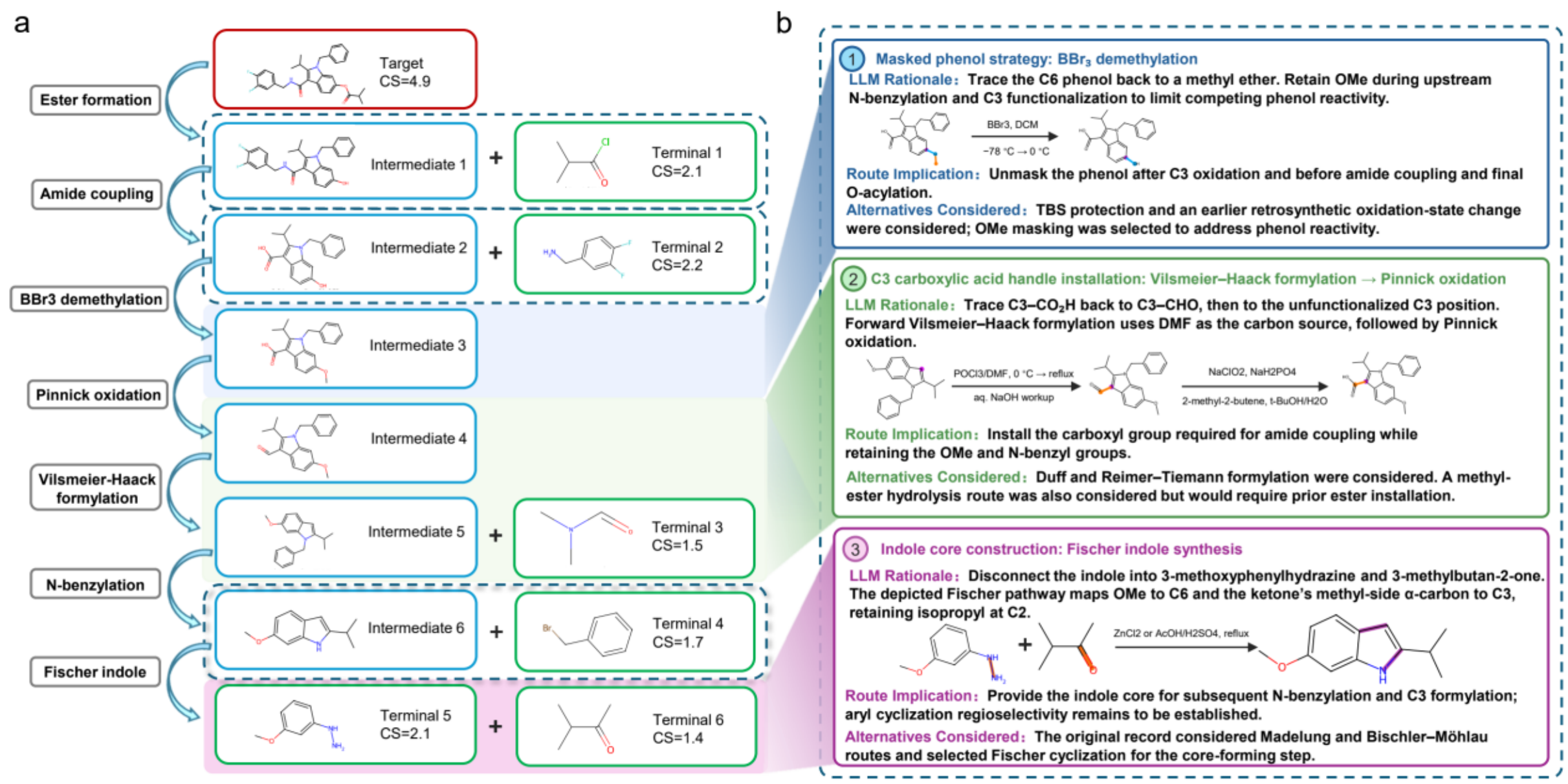

**Fig. 2 | LLM-guided retrosynthetic route construction for target n1_366. a,** A seven-step route connecting the target to six terminal precursors. The route is displayed in the retrosynthetic direction, whereas reaction labels denote the corresponding forward transformations: ester formation, amide coupling, $BBr_3$ demethylation, Pinnick oxidation, Vilsmeier–Haack formylation, N-benzylation and Fischer indole synthesis. Molecular complexity scores (CS) are shown for the target and terminal precursors. **b,** Three representative decisions illustrate how local reaction choices shape the overall route: masking the phenol as a methyl ether during upstream transformations; installing the C3 carboxyl group through formylation followed by oxidation; and constructing the indole core through Fischer cyclization. Reaction schemes are shown in the forward direction, accompanied by the **LLM rationale**, **route implication** and **alternatives considered**.

## Results

### A general-purpose LLM constructs complete routes within Rachel

We first asked whether a general-purpose LLM could construct complete retrosynthetic routes through successive chemical decisions. GPT-5.5 planned within Rachel from target structures, without a supplied reference route or route-level solution. After planning ended, each route was frozen and its terminal precursors underwent an independent source-availability audit using public-record evidence that had not been provided during planning. A route was classified as strictly closed only if construction was complete and every terminal met the source-resolution criterion; all other outcomes were classified as unresolved (Methods).

On PaRoutes120, Rachel reached strict closure for 111 of 120 targets, compared with 95 each for direct LLM and AOT* and 83 for SyntheLite (Fig. 3a). The separately analysed RF25 difficult-target cohort was drawn largely from studies published after GPT-5.5's reported knowledge cutoff. Here, Rachel reached strict closure for 24 of 25 targets, compared with 15 for direct LLM, 14 for AOT* and 6 for SyntheLite (Fig. 3b). Terminal acceptance was not determined by a scalar complexity cutoff alone: over a quarter of the final PaRoutes terminal nodes exceeded their run-specific thresholds (Figure S5). An accompanying Route Atlas links these outcomes to the available method routes for every target, retaining incomplete and unavailable outcomes and including the PaRoutes reference routes for comparison (Supplementary Data). We then assessed the chemistry connecting these endpoints to the targets at the reaction and whole-route levels.

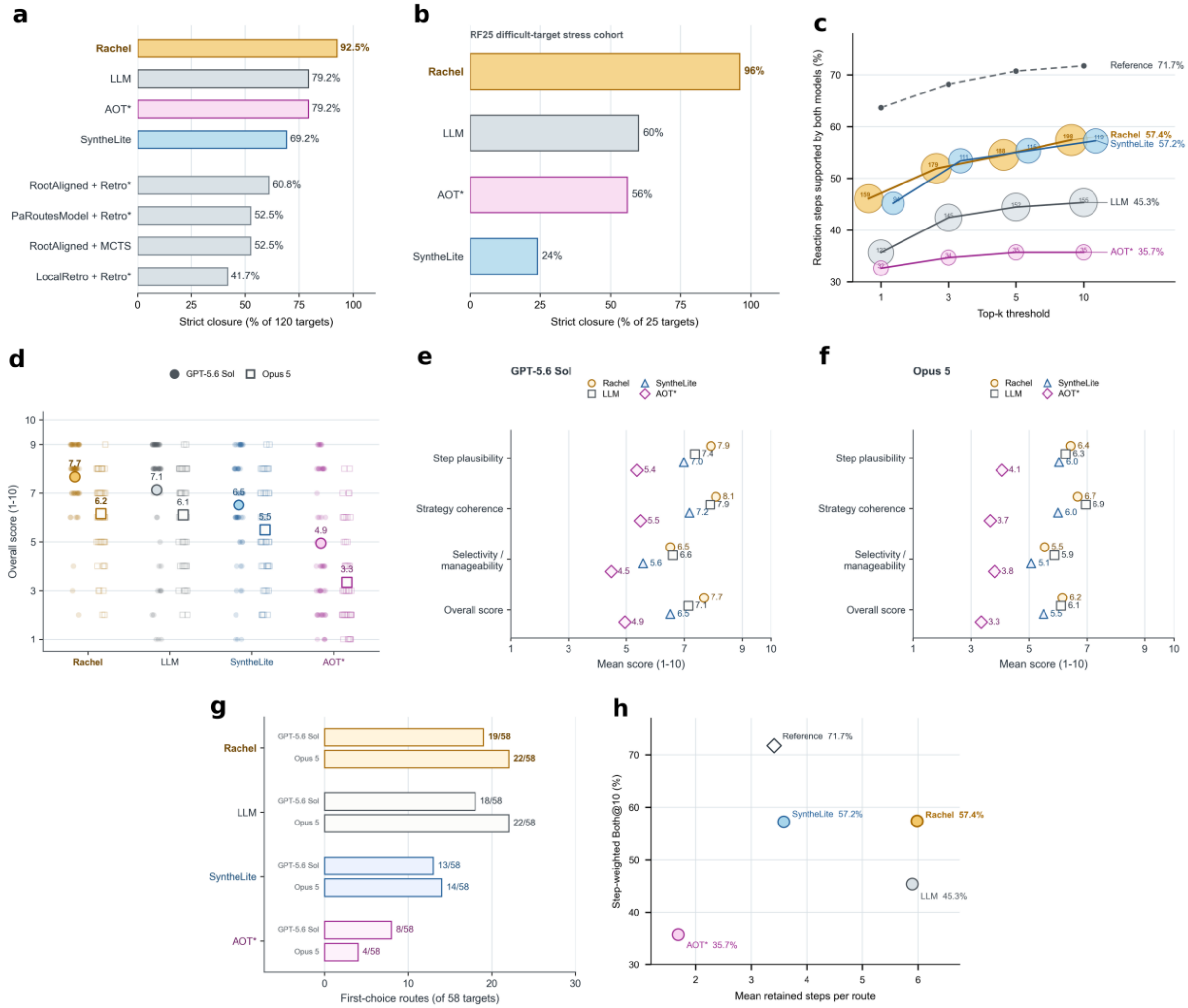

**Fig. 3 | Population-scale multistep route construction and complementary model-based chemical-quality assessment. a,** Strict-closure rates on the 120-target PaRoutes benchmark. Strict closure requires completed route construction and source resolution of every terminal (Methods). LLM denotes the direct GPT-5.5 comparator using task-specific structured prompts and structured-output requirements, without the Rachel stateful route framework. **b,** Strict-closure rates on the separate 25-target RF25 difficult-target cohort. The LLM definition in a applies. **c,** Step-weighted dual-forward support on common58, the 58 targets strictly closed by all four route-construction methods. Support fractions use steps successfully scored by both models. Both@k requires both ReactionT5v2-forward and RXNGraphormer to rank the recorded product within Top-k; the reference trajectory is shown separately. **d,** Method-blinded whole-route overall-score distributions on common58, with evaluator-specific means shown on the registered 1-10 scale. **e,** GPT-5.6 Sol scores across four dimensions of whole-route appraisal for 58 common58 targets. **f,** Opus 5 mean scores across four dimensions of whole-route appraisal for 58 common58 targets. For each target, four routes were presented in randomized A–D order without method labels, using only SMILES-based route representations; each mean is calculated from 58 integer scores on the 1–10 scale. **g,** Evaluator-specific first-choice route counts across 58 common58 targets. GPT-5.6 Sol counts select Rachel in Rachel-containing top ties; Opus 5 counts retain co-first ties. **h,** Mean retained steps per route versus step-weighted Both@10 on common58. Equal-area circles denote generated methods; the open diamond denotes the official reference. The display is descriptive and not adjusted for route length. Target-level Both@10 comparisons supplement panel c (Figure S9).

RF25 forward-model agreement is reported in Figure S19. Terminal complexity and conditional route morphology are shown in Figures S5, S6; complete advancement decisions and CS-BERTz profiles are shown in Figures S7, S8. Protocols and analysis rules are given in Supplementary Methods S2-S4 and S6.

## Broad route coverage coexists with computationally supported chemistry

Reaction-level support was close to SyntheLite's on common58, the 58 PaRoutes targets with strictly closed routes from Rachel, direct LLM, AOT* and SyntheLite. Each method generated its own route for each shared target. Both forward models recovered the recorded product within their top ten predictions for 57.4% of Rachel's jointly scored steps (Both@10), compared with 71.7% for the official PaRoutes reference routes. Rachel's step-weighted support exceeded that of direct LLM and AOT* (Fig. 3c). The similar support fractions accompanied more retained reaction steps per route for Rachel than for SyntheLite (Fig. 3h).

The whole-route assessment asked whether the proposed transformations formed a coherent synthesis strategy. In separate, method-blinded appraisals of standardized routes for the same targets, GPT-5.6 Sol and Opus 5 both assigned Rachel the highest mean overall score, although Opus 5 rated it closely to direct LLM (Fig. 3d). Both evaluators assigned Rachel slightly lower selectivity/manageability scores than direct LLM (Fig. 3e,f). Rachel and direct LLM also received similar numbers of first choices under the evaluator-specific tie rules (Fig. 3g).

For RF25, forward-model agreement was examined separately on each method's own strictly closed routes. Across Rachel's 24 routes, both models recovered the recorded product within their top ten predictions for 67 of 308 steps (21.8%; Figure S19), the largest supported-step count among the method-specific route sets, although not the highest fraction. To examine how the routes took shape, we next traced the recorded chemical choices and plan revisions.

## Chemical choices and route strategy evolve during construction

The route records showed how successive chemical choices entered a developing route. For the indole target, the LLM connected phenol masking, C3 functionalization and core construction into a complete route (Fig. 2). In the full-cohort analyses, including unresolved routes, model-proposed transformations outside Rachel's predefined operations appeared from early to late relative route positions in both PaRoutes120 and RF25. They formed the majority of committed chemistry throughout the RF25 position profile (Fig. 4a).

Alongside these successive chemical choices, the overall route plans were revised. At least one revision was recorded in 35 of 120 PaRoutes routes and 23 of 25 RF25 routes (Fig. 4b). The planning context recorded for each committed step connected these updates to subsequent chemistry: nearly every route with a recorded revision contained at least one retained transformation registered under an updated plan (Fig. 4c).

The fused heteroaryl target n5_7967 illustrates what such a revision changed chemically. Its initial plan retained the complex core without specifying how to assemble it. Before revision, reverse tracing through an ester-hydrolysis step had already reached the ester precursor. The revised plan specified construction of this ester by annulation of a substituted 2-aminopyridine with ethyl bromopyruvate. That annulation was subsequently incorporated into the route, defining the two precursors to be considered next (Fig. 4e).

The model also selected among proposals whose validation evidence remained incomplete. In both cohorts, such transformations entered final routes, while hard-blocked candidates remained excluded (Fig. 4d). In one local pair, the selected proposal supplied a methylene source absent from the unselected proposal (Fig. 4f). In the complete RF25 route, a selected proposal specified carbon tetrabromide and triphenylphosphine, whereas an unselected alternative omitted the bromine source; both carried the same missing-evidence status (Fig. 4g). We next examined whether complete-route construction was maintained when the LLM's route decisions or their supporting functions were altered.

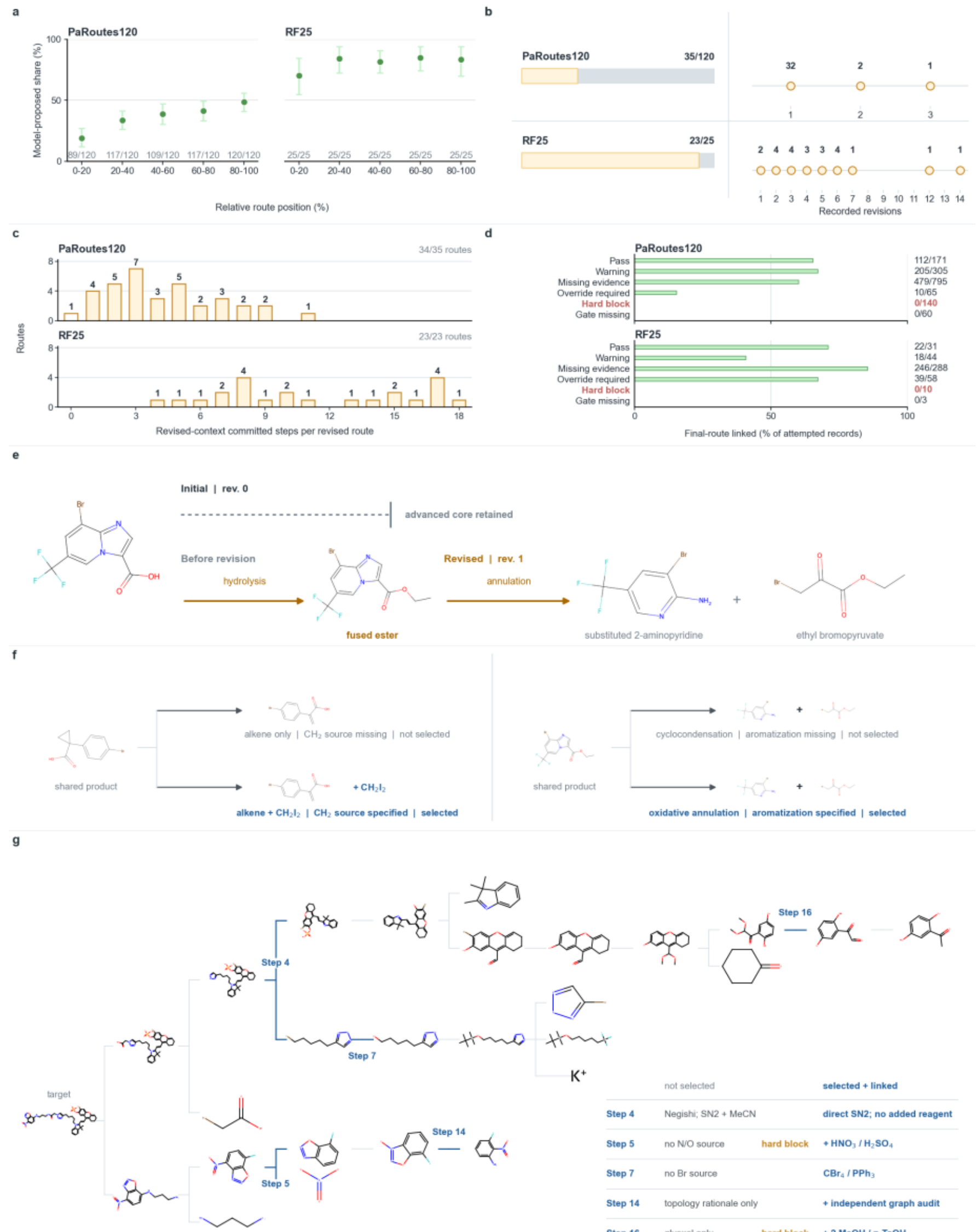

**Fig. 4 | Recorded chemical decisions and plan revisions during route construction. a,** Model-proposed chemistry across five equal-width route-position bins in PaRoutes120 and RF25, including unresolved routes. Position is committed reaction order divided by total route reactions. Green points show equally weighted means of within-target custom-transformation shares; bars show pointwise 95% intervals from 10,000 bootstrap resamples of contributing targets. Labels give contributing/registered targets; targets without a reaction in a bin are excluded from its mean. **b,** Fraction of registered routes with a recorded plan revision (left) and revision-count distribution among revised routes (right); point labels give route counts. **c,** Committed steps registered under revised plans, among revised routes. Bars and labels count routes; upper-right fractions identify routes containing such steps (PaRoutes120, 34/35; RF25, 23/23). **d,** Final-route linkage of attempted candidates across six overlapping validation or gate states. Labels give linked/attempted records. Missing evidence indicates an evidence limitation; Gate missing indicates an absent gate record. Of 247 selected RF25 Missing evidence records, 246 were linked to the final route. **e,** Plan revision for PaRoutes120 target n5_7967. Retrosynthetic tracing had reached the ester precursor before revision; the revised plan specified its construction by annulation of a substituted 2-aminopyridine with ethyl bromopyruvate, subsequently incorporated into the route. Reaction names denote forward transformations. **f,** Paired candidates contrast atom-source specification (left) and reaction-event specification (right), with recorded selection outcomes. **g,** Complete retrosynthetic tree for RF25 target rf25_008: 27 molecular nodes and 17 reactions. Records at Steps 4, 5, 7, 14 and 16 compare unselected candidates with those selected and linked to the final route; step numbers follow registered reaction order. These are records of planning and route incorporation, not experimental validation or isolated component effects. Supporting audits and route records appear in Figures S1, S2 and S10-S15; definitions and extraction rules are in Supplementary Methods S2 and S6.

Replacing route decisions or restricting planning support reduces complete-route construction

Replacing the LLM's route decisions separated local execution from complete-route construction. On the same 120 PaRoutes targets, three fixed policies continued to produce locally validated actions and extend route trees within Rachel, yet strict closure fell from 111/120 to 6–15/120 (Fig. 5a,b). These replacement experiments support a role for LLM-directed chemical choices in connecting local steps into complete routes.

We next conducted system-level component ablations on RF25, retaining the LLM as planner to examine route completion under restrictions on planning support (Methods). On the same 25 RF25 targets, restrictions on custom chemistry, persistent plans, route sketches/continuation or detailed validation feedback each reduced strict closure, from 24/25 to 2–18/25 (Fig. 5c).

The reduced closure was accompanied by changes in the chemistry incorporated into routes. When custom proposals were disabled, committed transformations came only from predefined bond-disconnection and functional-group operations. Under the other restrictions, model-proposed transformations still appeared in most routes, but fewer were incorporated overall, alongside more predefined operations than under native Rachel (Fig. 5d).

RF25 target 014 shows how these differences unfolded within a single molecular problem. The four restricted runs produced different route trees without reaching strict closure, with some routes sharing terminal molecules with native Rachel. Native Rachel connected the target to source-resolved terminals, completing the route (Fig. 5e).

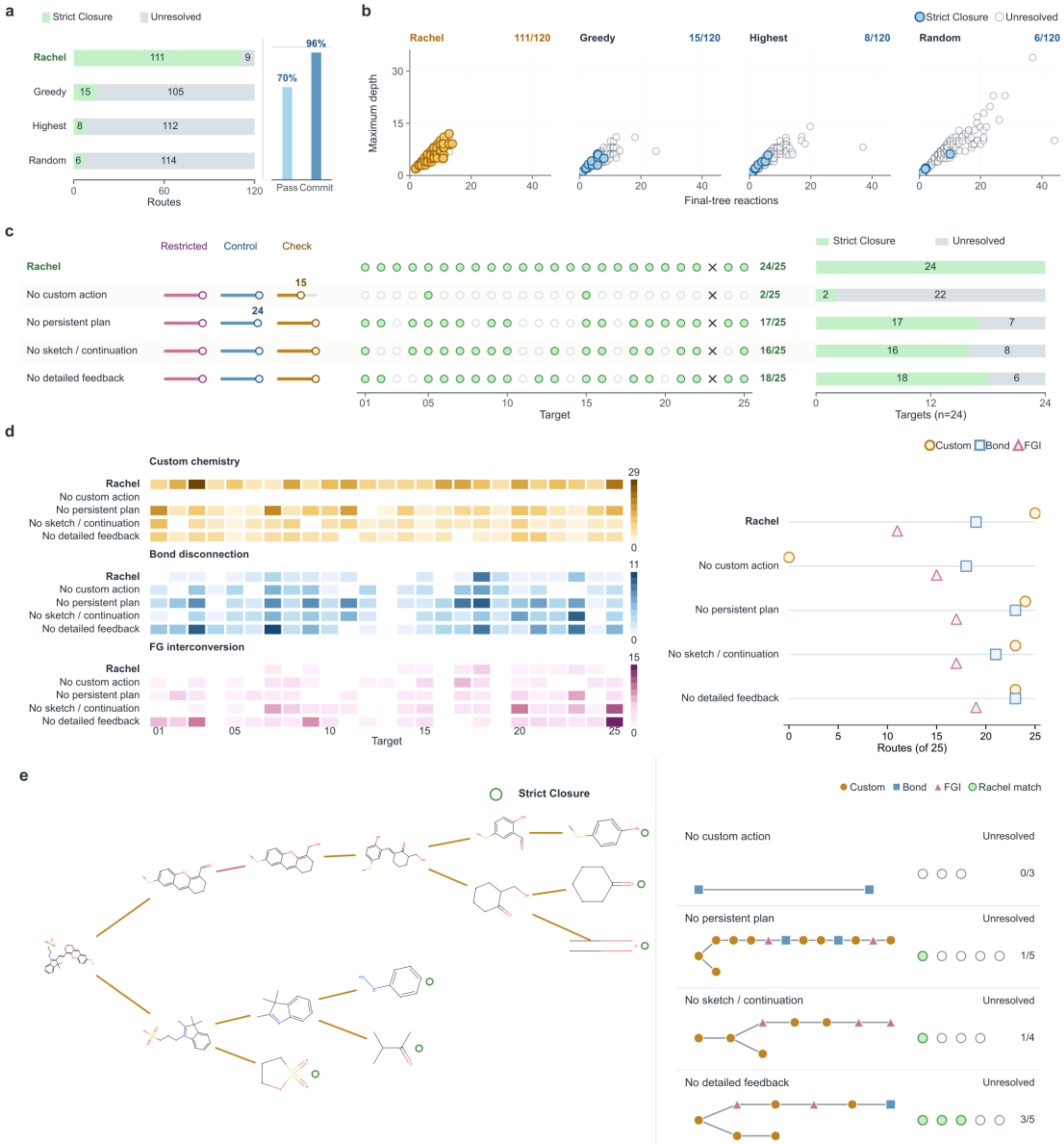

**Fig. 5 | Route construction under decision replacement and planning-support restrictions. a,** Strict Closure and Unresolved outcomes for native Rachel and fixed mechanical policies on the same 120 PaRoutes targets. Pooled mechanical audits report passed candidates (22,577/32,295, 70%) and committed decision opportunities (2,894/3,016, 96%), with distinct denominators. **b,** Final-tree reaction count versus maximum depth per target-method pair; filled/open points denote Strict Closure/Unresolved. These are descriptive route measures. **c,** System-component ablations on the same 25 RF25 targets: no custom action, no persistent plan, no sketch/continuation and no detailed model-visible feedback. The last condition retains private validation and the commit gate. Left: target counts out of 25 with the intended signal absent (Restricted), retained-function use recorded (Control), or an execution check recorded (Check); condition-specific definitions are in Figure S4. Middle: outcomes for all 25 targets. Filled green circles denote closure; open circles mark Rachel-closed targets unresolved under restriction; crosses mark target 023, unresolved throughout. Right: binary outcomes restricted to the 24 native-Rachel-closed targets. **d,** Target-level counts of committed Custom, Bond and functional-group interconversion (FGI) actions, with a separate heatmap scale for each source. The right-hand summary counts routes containing at least one action from each source (circle, Custom; square, Bond; triangle, FGI). **e,** Target 014: Rachel's molecular route (left) and restricted-route topologies and outcomes (right). Green terminal marks denote exact canonical-SMILES matches to Rachel terminal leaves. Fractions give matching/all terminal leaves (0/3, 1/5, 1/4 and 3/5, in row order); a terminal match alone does not establish route closure. PaRoutes120 and RF25 are separate experiments. Strict closure requires complete construction and independent source resolution of every terminal, not laboratory synthesis. Supporting audits, morphology and terminal identities are in Figures S3, S4, S16 and S17; protocols and analysis rules are in Supplementary Methods S3, S5 and S6.

## Discussion and Conclusion

Each accepted retrosynthetic step introduces precursor molecules for the planner to address. Within Rachel, a general-purpose LLM takes up these successive problems and revises its route strategy as planning proceeds. Across the tested targets, the model often constructed routes whose terminal precursors subsequently passed an independent source audit. Fixed policies could still execute local chemistry but rarely reproduced this outcome; systems with restricted planning support also completed fewer routes. These system-level observations support a role for general-purpose models in connecting chemical decisions across a route. The route records allow local choices to be studied alongside revisions to the overall synthesis strategy.

Turning a computed route into a synthesis requires conditions under which its key transformations work.[21] Rachel's RDKit-based structure checks and reaction rules provide feedback on encoded constraints, leaving reactivity under specific conditions to be established. For stereochemically demanding targets, the route strategy also depends on how individual transformations establish the required configurations. An accepted error can change subsequent molecular problems, so independent chemical evidence is needed throughout planning. Model-proposed precursor sets extend the available chemistry beyond predefined operations, but their incorporation into a route carries the same need for independent assessment. Post hoc forward predictions offer a separate check. Agreement was lower in RF25 than in common58 and is reported as a supplementary diagnostic (Figure S19). Human synthetic-expert assessment of these difficult-target routes remains to be undertaken, focusing on key transformations, selectivity and compatibility. Future planners could use more discriminating calculations or experiments to revisit consequential choices, while retaining alternative routes when the available evidence does not favour a single strategy. Whether richer evidence improves these decisions can be tested through expert review and prospective synthesis, together with repeated runs across model families. Rachel makes it possible to trace how such feedback changes both the current choice and the chemical problems encountered next.

Beyond retrosynthesis, this approach suggests how large models might reshape the connections between specialized chemical tasks. Molecular design could respond to difficulties encountered in synthesis: a molecule selected for its predicted properties but difficult to prepare could be redesigned to retain the desired function.[22,23] An unexpected experimental product might, in turn, prompt a revised mechanistic hypothesis and a different next experiment. Such coordination would require the model to carry evidence from synthesis and measurement into subsequent design. When observations conflict with earlier assumptions, it would also need to identify what to investigate next to resolve the discrepancy. Rachel allows one part of this problem to be studied in computational planning, where accepted transformations change the molecular problems that follow.

The longer-term prospect is for general-purpose models to help organize an entire chemical investigation. As their capabilities improve, models may maintain and revise plans with less external support. Execution records and independent verification could still keep those decisions open to scrutiny. Specialized predictors, molecular simulations and experimental platforms could supply evidence as needed, while chemists retain control of research objectives and experimental execution.[24,25] In catalyst development or molecular materials discovery, a finding about reactivity or function could change both the candidate structures and the experiments used to examine them. If models can reliably sustain this exchange between evidence and decisions, they could help chemists decide how a study should proceed as new findings emerge.

## Computational Methods

### Target sets

We evaluated LLM-directed multistep retrosynthetic planning with Rachel in a computational study; no laboratory synthesis was performed. PaRoutes120 comprised 120 targets from the PaRoutes dataset[7], including 51 n1 and 69 n5 targets.

RF25 comprised 25 targets documented in molecular-probe and chemical-biology studies and was analysed separately. Source studies for 24 targets were published in 2026, after GPT-5.5's reported knowledge cutoff of 1 December 2025. The remaining target came from a study published in September 2025. Molecular weights ranged from 231.3 to 2082.4 Da, and heavy-atom counts ranged from 16 to 149. Target identifiers, input SMILES, source references and molecular descriptors are provided in the Supplementary Data.

### Retrosynthetic planning protocol

GPT-5.5, with reasoning effort set to High, constructed routes through Rachel-beta from a target structure and an initial task instruction, without a supplied reference route, route-level solution or externally written synthetic plan. Context relevant to the active molecule, candidate operations, validation results and current route strategy was progressively disclosed as the session state changed. An initialization example and the context-update procedure are provided in Supplementary Methods S2 and Table S1.

For each active molecule, the LLM proposed or selected a transformation, decided whether to accept it, and chose whether to continue decomposition or accept the molecule as terminal. Rachel executed the operations and recorded the decisions. Accepted transformations were committed to the route, and intermediates requiring further decomposition entered the pending queue. A persistent route plan recorded the global strategy, whereas a local route sketch recorded a proposed continuation. Inspection and candidate attempts did not themselves add reactions to the route.

Transformations used predefined chemical operations or model-proposed precursor sets. Rachel used RDKit-based structural and reaction-rule checks as a lightweight, demonstration-level planning verifier. These checks assessed encoded constraints and informed the LLM's acceptance decision, but did not establish experimental reaction feasibility. Hard blocks prevented commitment, and designated overrides required a recorded justification. Molecular-complexity scores provided advisory context for terminal decisions. The built-in checks are described in the Supplementary Methods; post hoc reaction-level evaluation is described below.

On PaRoutes120, Rachel was compared with direct LLM, AOT*[16], SyntheLite[17], PaRoutesModel + Retro*[5,7], RootAligned + Retro*[5,26], RootAligned + MCTS[26] and LocalRetro + Retro*[5,27]. RF25 comparisons used direct LLM, AOT* and SyntheLite. Available experiment-specific model configurations, execution settings, budgets and stopping rules are indexed in the Supplementary Methods and accompanying data.

## Route closure assessment

Internal stopping and strict route closure were assessed separately. After planning ended and the route was frozen, terminal structures were canonicalized and deduplicated with RDKit[28] and audited against PubChem compound identifiers (CIDs) and Chemical Vendors records.[29] A terminal was source-resolved when both a CID and vendor evidence were available. Registered in situ metal reagents could instead be resolved through an approved preparation-source mapping, provided that the specified source form or every required source reagent met the same CID and vendor-evidence criterion.

The same closure criterion was applied across methods in PaRoutes120 and RF25. A route was classified as Strict Closure only if construction was complete and every terminal was source-resolved; all other outcomes were classified as Unresolved. Every target remained in the corresponding benchmark denominator, including those with missing or failed route outputs. Terminal inventories and the CID, vendor and preparation-source records underlying these decisions are provided in the Supplementary Data. Audit information was not available during planning and was not fed back into route construction. Strict closure denotes this computational source-resolution endpoint and does not establish synthetic feasibility or confirmed procurement.

Reaction and route evaluation

Matched-target quality comparisons used common58, the 58 PaRoutes120 targets for which Rachel, direct LLM, AOT* and SyntheLite all produced strictly closed routes (24 n1 and 34 n5 targets). Each method contributed its independently generated route for each target; reaction steps were not matched across methods. Corresponding PaRoutes reference routes provided the reference comparison. RF25 reaction-level evaluation used each method's own strictly closed routes, so the target subsets could differ between methods.

Reaction-level support was assessed with ReactionT5v2-forward[30] and RXNGraphormer[31], using the sagawa/ReactionT5v2-forward checkpoint and the official USPTO-480k valid_checkpoint.pt, respectively. Recorded reactants were supplied without reaction conditions. Each model used a beam size of 10 and returned up to ten ranked product predictions. Product recovery was scored at the first matching prediction. Fixed checkpoint identifiers and inference settings are documented in Supplementary Methods S4 and Table S3.

Top-k support was defined as recovery of the recorded product within the first k predictions (k = 1, 3, 5 and 10). Both@k required both models to meet this criterion for the same step. Within each route-construction method, all support fractions used the steps successfully scored by both models, pooled with equal weight per step. Two of Rachel's 347 retained common58 steps could not be scored by RXNGraphormer, leaving 345 steps in the joint denominator. Scoring coverage and preprocessing details are recorded in the Supplementary Methods and Data. Both@k measures agreement with the recorded product under this prediction protocol, not the fraction of experimentally feasible reactions.

Whole-route appraisal on common58 was performed separately by GPT-5.6 Sol and Opus 5. For each target, each evaluator received the target structure and four frozen routes in a single request, with routes represented by terminal SMILES and forward-ordered reaction steps. Method identities were concealed behind labels A-D, with balanced assignments and the same mapping for both evaluators. Each route was scored from 1 to 10 for step chemical plausibility, route strategy coherence, selectivity and compatibility manageability, and overall advancement to human synthetic review, accompanied by a yes, uncertain or no recommendation. Scores and rankings were summarized separately for each evaluator. The review instructions, scoring rubric and label assignments are documented in Supplementary Methods S4 and Table S4. These were model appraisals; human-expert assessment of RF25 routes was not performed.

## Control and ablation experiments

On PaRoutes120, LLM route-action decisions were replaced with three mechanical policies operating on successfully executed, validation-passed candidates within Rachel. Greedy complexity maximized the difference between the current molecule's complexity score and the highest precursor score. Highest score selected the candidate with the highest heuristic score, and Random valid sampled uniformly from eligible candidates. The policies used Rachel's candidate execution and validation procedures, and route closure was assessed on all 120 targets using the criterion above. Candidate enumeration, tie-breaking, randomization and execution settings are documented in Supplementary Methods S5 and Table S5.

System-component ablations on RF25 separately removed the custom-chemistry channel, persistent route plan, local route sketch and continuation, or detailed validation feedback. The feedback ablation retained the model-visible gate status, internal validation and commitment constraints. Each ablated condition was run through the API with GPT-5.5 at High reasoning effort, with one route per target-condition and a budget of 200 turns. Component removal was checked against recorded requests and session events. Strict closure was compared with native Rachel across all 25 targets; conditional route-state analyses used the 24 targets strictly closed by native Rachel. Native and ablated conditions retained their recorded execution configurations, and comparisons were interpreted as system-level contrasts. Full component definitions, run configurations and implementation checks are provided in Supplementary Methods S5, Tables S6, S7 and Figure S4.

## Data analysis

Process analyses used the recorded sessions and frozen routes for all PaRoutes120 and RF25 targets, including unresolved routes, and were summarized separately by cohort. Plan revisions were identified from recorded version histories and linked to retained reaction steps through their plan-version references. Missing history was retained as unavailable in the extraction tables. Candidate attempts, selections and final-route incorporation were distinguished using the corresponding event and reaction identifiers. Detailed extraction rules, route-size definitions, case-selection procedures and terminal-identity matching are documented in Supplementary Methods S6 and Table S8.

For the route-position analysis, each retained reaction's position was its order in the committed sequence divided by the route's total reaction count. Within each of five equal-width position bins, the proportion of model-proposed custom transformations was calculated for each target and then averaged with equal weight across contributing targets. Targets with no reaction in a bin were excluded from that bin's mean. Contributing targets were resampled with replacement 10,000 times, separately for each cohort and bin. The 2.5th and 97.5th percentiles of the resulting bootstrap means defined pointwise 95% intervals.

## Associated Content

Supporting Information is appended after the main references and contains computational protocols, evaluation definitions, supporting figures and tables. The PaRoutes120 and RF25 Route Atlas (HTML) and Source Data with supporting documentation (ZIP) are available through the public links below.

## Data Availability

The Rachel Route Atlas is available on GitHub, with a fixed repository version 57c1716. It provides offline route comparisons for all 120 PaRoutes targets and 25 RF25 targets, including separate PaRoutes reference routes. Figure-level data, machine-readable records and supporting documentation are available in the Source Data release, version 2026-09-07. Rachel-authored materials are distributed under CC BY-NC 4.0; third-party terms are retained.

## Code Availability

Rachel code, installation instructions and user guidance are available at https://github.com/ChazenLi/Rachel, with the present release identified by repository version 57c1716. The repository license is CC BY-NC 4.0; third-party dependencies retain their own terms. Experimental protocols and recorded configurations are described in Methods and Supporting Information.

**Author Information**

Corresponding Authors: Xin Su: xinsu@mail.buct.edu.cn; Da Han: dahan@sjtu.edu.cn; Guangyong Chen: gychen@link.cuhk.edu.hk.

# Supporting Information

Rachel: A general-purpose language model directs and revises retrosynthetic routes

Qisheng Li[1,2], Shunchao Jiang[3], Chen Qi[1], Xin Su[4,*], Da Han[5,6,*], Guangyong Chen[1,*]

[1] Hangzhou Institute of Medicine, Chinese Academy of Sciences, Hangzhou, China

[2] Beijing Advanced Innovation Center for Soft Matter Science and Engineering and State Key Laboratory of Organic–Inorganic Composites, Beijing Key Laboratory of Bioprocess, College of Life Science and Technology, Beijing University of Chemical Technology, Beijing, China

[3] Zhenzhida Biotechnology (Shanghai) Co., Ltd.

[4] State Key Laboratory of Organic–Inorganic Composites, Beijing Advanced Innovation Center for Soft Matter Science and Engineering, College of Life Science and Technology, Beijing University of Chemical Technology, Beijing 100029, China

[5] Zhejiang Key Laboratory of Functional Nucleic Acids for Basic and Clinical Application, Hangzhou Institute of Medicine, Chinese Academy of Sciences, Hangzhou, China

[6] Institute of Molecular Medicine, Renji Hospital, School of Medicine, Shanghai Jiao Tong University, Shanghai, China

* Correspondence: Xin Su: xinsu@mail.buct.edu.cn; Da Han: dahan@sjtu.edu.cn; Guangyong Chen: gychen@link.cuhk.edu.hk.

Supplementary methods, parameter and definition tables, and supporting figures for the Rachel study.

## Contents

## S1. Target sets

### S1.1 Cohort definition

PaRoutes120 comprised 120 targets from the PaRoutes dataset[1], including 51 n1 and 69 n5 targets. The fixed target inventory supplied the identifiers used to link input structures, generated routes and endpoint records across methods. All 120 targets remained in each method's closure denominator.

RF25 comprised 25 targets documented in molecular-probe and chemical-biology studies and was analysed separately. The source records included 24 targets reported in 2026 and one reported in 2025. Molecular descriptors were calculated from the recorded input structures with RDKit.[2] Molecular weights ranged from 231.3 to 2082.4 Da, with 16-149 heavy atoms.

### S1.2 Minimum target-record fields

The Supplementary Data retain one stable identifier for each target, the input structure used for computation, cohort membership and the provenance pointer. For RF25, the registry additionally retains the source publication and the molecular descriptors reported above. These fields support target matching and audit; they are not a difficulty score or a reconstruction of historical sampling.

### S1.3 Supplementary-data location

The target inventory, RF25 registry and descriptor records are provided as Supplementary Data. GPT-5.5 route outputs are linked through the target and route identifiers. Complete target rows and descriptors are supplied as machine-readable files rather than reproduced in this section.

## S2. Retrosynthetic planning protocol

Rachel planning used GPT-5.5 with reasoning effort set to High, operating through the Rachel-beta harness. A target structure and an initial task instruction initiated the session. During planning, context relevant to the active molecule, candidate operations, validation results and current route strategy was progressively disclosed as the session state changed. The authors did not specify an additional temperature setting for these Rachel runs. Available experiment-specific settings are indexed in the accompanying data package.

The LLM proposed or selected transformations, decided whether to accept them, and chose whether to continue decomposition or accept a terminal. Rachel retained the active molecule, pending intermediates, candidate attempts and accepted route in a persistent session. The route plan recorded the global strategy, whereas the route sketch recorded a proposed local continuation. Inspection and candidate attempts did not themselves add reactions to the route. Accepted transformations were incorporated through commitment. Validation feedback informed acceptance; hard blocks prevented commitment, and designated overrides required a recorded justification. RDKit-based structure checks and reaction rules assessed encoded constraints and supplied planning-time feedback. Acceptance of a predefined or model-proposed transformation did not establish its feasibility under experimental conditions. This feedback was distinct from the post hoc forward-model evaluation in S4.

The initialization example below is reproduced in its original Chinese form with an English translation. Model-visible context also included runtime instructions and subsequent tool responses. It therefore represents the starting instruction, not the complete context of a planning session.

**Initialization example: original**

新建独立运行目录，按 `Rachel` 的状态化工作流对下面的分子进行高质量、真实的从头逆合成，大胆尝试、设想设计、严格验证。每次提交仅包含一个真实化学事件；在 `commit` 前核查机制、原子来源、位点与拓扑保真、官能团兼容性、选择性和立体化学：`SMILES: O=C1Nc2ccc(F)cc2C12C(C)CCCC2`

**Initialization example: English translation**

```
Create a separate run directory and use Rachel's stateful workflow to develop a high-quality,
chemically realistic de novo retrosynthesis for the molecule below. Explore boldly, propose designs
and validate rigorously. Each commit must contain only one actual chemical event. Before committing,
check the mechanism, atom sources, site and topology fidelity, functional-group compatibility,
selectivity and stereochemistry. SMILES: O=C1Nc2ccc(F)cc2C12C(C)CCCC2
```

### S2.1 Operational state fields

**Table S1 | Operational state fields.**

| Recorded object | Operational meaning |
|---|---|
| Active molecule and queue | Current decomposition target and intermediates awaiting further decisions |
| `route_plan` | Persisted global strategy and its revisions |
| `route_sketch` | Proposed local continuation |
| Candidate attempts and validation | Candidate chemistry, returned evidence and admissibility state |
| Commit and accepted route | Transformations incorporated into the route |
| Tool-return context | State- and event-relevant planning summaries, feedback and available next operations |

Available experiment manifests and comparator-specific settings are indexed in the accompanying data package.

Rachel used RDKit-based chemical checks as a lightweight, demonstration-level planning verifier, including atom balance, template execution, scaffold and topology checks, and functional-group compatibility. These checks constrain encoded structures and transformations but do not establish experimental feasibility. The benchmark runtime implementation and configuration records accompany the data.

## S3. Route closure assessment

Terminal structures were extracted from the frozen route, canonicalized and deduplicated with RDKit, and linked to archived PubChem CID and Chemical Vendors records.[3] A terminal was source-resolved when both a CID and vendor evidence were available. Registered in situ metal reagents could be resolved through a preparation-source mapping if the designated source form or every required source reagent met the same criterion. These records establish the declared source-evidence endpoint at the time of the audit.

Strict Closure required a completed route with every terminal source-resolved. All other outcomes were Unresolved, including missing or failed outputs. Internal stopping was recorded separately from this endpoint. The route-level decision was reconstructed by joining the full terminal inventory to the query records and registered mappings, retaining unresolved terminals in the inventory. The post hoc audit was conducted after route freezing and was not fed back into planning.

### S3.1 Minimum terminal-evidence fields

**Table S2 | Minimum terminal-evidence fields.**

| Record group | Minimum fields retained in Supplementary Data |
|---|---|
| Route identity | Cohort, method, target ID, frozen route ID or source pointer |
| Terminal identity | Recorded structure, canonical structure and route linkage |
| External evidence | CID, vendor evidence and retained query/cache/source reference |
| Preparation mapping | Registered terminal-to-source mapping and evidence for required source forms/reagents |
| Outcome | Terminal source resolution and route Strict Closure / Unresolved |

The complete terminal-by-route evidence table, including CID, vendor records, source mappings and unresolved entries, is supplied as Supplementary Data. It is not reproduced in the Supplementary Methods.

## S4. Reaction and route evaluation

The common58 comparison used the 58 PaRoutes120 targets strictly closed by Rachel, direct LLM, AOT* and SyntheLite, comprising 24 n1 and 34 n5 targets. Each method contributed one frozen route per target. Corresponding PaRoutes reference routes provided the reference comparison. RF25 forward evaluation used each method's own strictly closed routes; its target subsets could therefore differ between methods.

ReactionT5v2-forward[4] and RXNGraphormer[5] predicted products from recorded reactants without supplied reaction conditions. The archived worker protocol and fixed checkpoints are listed below. RXNGraphormer input preprocessing wrote non-isomeric reactant SMILES. Product scoring converted both predicted and recorded products to canonical isomeric SMILES and retained the first exact match at its original one-based rank. Invalid predictions were skipped without renumbering later predictions; duplicate predictions were retained.

For k=1, 3, 5 and 10, model-specific Top-k support was the fraction of evaluable steps whose recorded product appeared within the first k predictions. Both@k required recovery by both models for the same step. Within each route-construction method, all fractions used the steps available from both workers, with equal weight per step. Coverage counts and unavailable-step records are supplied as machine-readable Source Data. Both@k records predictive agreement under this protocol, rather than experimental correctness. A nonmatch can reflect an error in the proposed chemistry or a limitation of the predictor and its inputs; the metric alone does not distinguish these explanations.

For reference-route extraction, each reaction's parent molecule was the recorded product and its immediate precursor molecule nodes were the reactants. Atom-map annotations were removed. Reagent and condition metadata were retained as provenance and excluded from forward-model inputs. The full reference archive contains 120 routes and 426 steps; the common58 subset contains 58 routes and 198 steps. Predictions and metadata were recovered from archived results produced on the evaluation machine.

Whole-route appraisal used one request per target containing four anonymous routes labelled A-D. Each route was represented by target and terminal SMILES and forward-ordered reaction steps. The method-to-label assignment was balanced using seed 2026072501; each method appeared under each label 14 or 15 times across the 58 targets. Both reviewers received the same mapping. GPT-5.6 Sol and Opus 5 scored the routes separately using the four dimensions below and provided a yes, uncertain or no recommendation for progression to human synthetic review. These recommendations were model judgements. Scores and rankings were summarized separately for each reviewer. No human-expert assessment of RF25 routes was performed in this study.

### S4.1 Forward-model configuration

**Table S3 | Forward-model configuration.**

| Item | ReactionT5v2-forward | RXNGraphormer |
|---|---|---|
| Checkpoint | sagawa/ReactionT5v2-forward | Official USPTO_480k/model/valid_checkpoint.pt |
| Fixed identifier | Revision 933114058cb2604dc1bf536dbebdfcefbe83d4fc | Source commit 0b5824328422b33a8c80ceb17576146f79b951ae |
| Input | REACTANT:<dot-separated reactants>REAGENT:; empty reagent field | Dot-separated reactants; non-isomeric serialization; neutral bracketed halides normalized for vocabulary compatibility |
| Input length | Maximum 150 tokens; truncation enabled | Vocabulary checked; unsupported tokens return unavailable status |
| Decoding | Beam 10; up to 10 products; maximum length 300; batch 8 | Beam 10; n_best 10; maximum length 300; batch 8; recorded inference temperature 1.0 |

The recorded RXNGraphormer temperature of 1.0 applies to forward-model inference and is separate from the LLM planning settings in S2.

Coverage counts and unavailable-step records for common58 are supplied as machine-readable Source Data. They describe the analysed sample, rather than defining an evaluation rule.

### S4.2 Whole-route appraisal rubric

**Table S4 | Whole-route appraisal rubric.**

| Dimension | Scoring question | Scale |
|---|---|---|
| Step chemical plausibility | Are the stated reactant-to-product changes chemically credible? | 1-10 |
| Route strategy coherence | Are intermediates, step order, convergence and synthetic logic coherent? | 1-10 |
| Selectivity and compatibility manageability | Are structural selectivity and functional-group risks manageable? | 1-10 |
| Overall advancement | Does the route merit progression to human synthetic review? | 1-10 |

The appraisal protocol required structure-based evaluation without external searches or attempts to infer route provenance. Missing conditions and yields were reflected as uncertainty in the scores and rationales rather than supplied by the reviewer.

Figure S19 reports the RF25 evaluation on each method's own strictly closed routes. Rachel, direct LLM, AOT* and SyntheLite contributed 24, 15, 14 and 6 routes, respectively, containing 308, 104, 65 and 33 jointly scored steps. All retained steps in these sets were scored by both models. Both@10 recovered 67/308 (21.8%), 19/104 (18.3%), 10/65 (15.4%) and 9/33 (27.3%) recorded products, respectively. Rachel contributes the largest absolute supported-step count in this comparison, while the route and step populations differ across methods; this count does not establish a higher support fraction or a matched-target quality ranking. Model-specific predictions and all k thresholds are retained in Source Data.

## S5. Control and ablation experiments

Mechanical policies operated on candidates successfully executed within Rachel whose recorded `forward_validation.pass` flag was true. This was the planning-time eligibility flag, distinct from the post hoc dual-forward scoring in S4. Bond alternatives were attempted before functional-group interconversions. Greedy complexity maximized the decrease from the current molecule's complexity score to the highest precursor score. Highest score maximized the recorded bond heuristic score, with a score of zero assigned to functional-group interconversions. Random valid selected uniformly from eligible candidates. The detailed ordering and randomization rules are listed below. If no candidate was eligible, the sandbox was cleared and the active molecule was skipped; route outcomes were subsequently classified using the shared closure criterion.

RF25 system-component ablations separately removed custom chemistry, persistent route planning, local sketch and continuation, or detailed validation feedback. Each condition used the API with GPT-5.5 at high reasoning effort and one route per target-condition. The formal budget was 200 turns, a maximum depth of 15 and 50 steps. Request retries and transport recovery continued the same session and were not counted as independent route repetitions. Component removal was checked against recorded requests and session events. Strict closure used all 25 targets, with native Rachel as reference. Conditional route-state comparisons used the 24 targets strictly closed by native Rachel.

### S5.1 Mechanical policy definitions

**Table S5 | Mechanical policy definitions.**

| Policy | Ordering and implementation |
|---|---|
| Greedy complexity | Descending `current_cs - max(precursor_cs_scores)`, then descending heuristic score, then ascending source rank |
| Highest score | Descending bond heuristic score (FGI=0), then descending complexity decrease, then ascending source rank |
| Random valid | Uniform `random.Random(...).choice`; base seed 20260402 plus route ID |
| Identical sort keys | Stable enumeration order |
| Eligibility | `attempt.success` and `forward_validation.pass is True` |

### S5.2 Component interventions

**Table S6 | Component interventions.**

| Condition | Removed | Retained |
|---|---|---|
| No custom action | Custom precursor/repair channel and exposed commands | Other chemistry operations, validation, plan/sketch and route records |
| No persistent plan | `route_plan` commands, model-visible fields and commit associations | Local sketch/continuation, chemistry, validation and route records |
| No sketch / continuation | Sketch and continuation commands/fields | Persistent plan, ordinary single-step chemistry, validation and route records |
| No detailed feedback | Detailed validation explanations in model-visible results | Gate state, internal validation and commitment constraints |

### S5.3 Additional API settings

**Table S7 | Additional API settings.**

| Setting | Recorded value |
| --- | --- |
| Maximum completion tokens | 4096 |
| Tool execution | Responses API native tools; one call per turn; persisted output items and tool returns replayed locally |
| Request timeout | 900 s |
| Attempts per request | 3 |
| Terminal CS setting | 2.25; planning setting, separate from post hoc Strict Closure |

## S6. Data analysis

Process analyses linked frozen route records to session histories using target, event, reaction and plan-version identifiers. A recorded plan revision had a revision index greater than zero. Retained steps were associated with revised plans through their plan-version references. Missing version history was retained as unavailable. Complete committed-step records are shown in Figure S10. Revision trajectories compare the literal active-node, key-disconnection, precursor-logic and route-thesis fields between successive versions (Figure S11). Three RF25 events with unresolved sources retain that status in Source Data; their white glyphs do not establish unchanged fields. The independent chemical revision ledger likewise counts exactly retained, added and removed entries (Figure S12).

For the route-position profile, reaction i in a retained committed sequence of N reactions was assigned position i/N. The sequence was divided into five equal-width bins. Within each target and bin, the custom-transformation fraction was the number of steps labelled `llm_proposed_custom` divided by all retained steps in that bin. Fractions were averaged with equal weight across contributing targets. A target with no step in a bin did not contribute to that bin's mean. Pointwise 95% intervals were the 2.5th and 97.5th percentiles of 10,000 bootstrap means, resampling contributing targets with replacement separately for each cohort and bin.

The revised-plan position profile used the same binning and target weighting, with revised-plan steps as the numerator (Figure S1). Available plan histories defined 104 eligible PaRoutes120 routes and 25 RF25 routes. The contributing-target counts across bins were 77, 103, 93, 103 and 104 for PaRoutes120, and 25 in every RF25 bin. This population differs from the 35 and 23 revised routes used in Fig. 4c.

Candidate attempts, event selections and final-route incorporation were distinguished by their recorded identifiers and flags. The validation-state inventories in Fig. 4d could overlap. For episodes containing a hard-block record, another candidate admitted to the final route and no selection were separate episode outcomes (Figure S2). The complete target audit reports episode counts in its first row and binary presence of these outcomes in the other two rows; both indicators can occur within one target across different episodes (Figure S13).

Inherited-state analyses counted each unique molecular-node identity once within its route. Identities were classified as active at the next commit, active after at least one intervening commit, or retained as final route endpoints. Identity-level compositions and route-level fractions were reported separately (Figure S14). Consecutive and delayed examples retain the recorded node identities and reaction order (Figure S15).

Supporting intervention audits preserve diagnostic categories within the binary route endpoint. Internally complete but audit-open routes, unresolved skips and budget-exhausted routes remain Unresolved (Figure S3 and Figures S16, S17). Figure S4 records use of the restricted signal, whereas Fig. 5c displays its absence; the component definitions are given in Table S6. The nonclosed morphology panel uses the 24 native-Rachel-closed targets, yielding 22, 7, 8 and 6 nonclosed routes under the four restrictions. The closed morphology panel uses each condition's strictly closed set: 24 native Rachel routes and 2, 17, 16 and 18 restricted routes (Figure S16).

Route reaction count and longest-path depth were treated as separate quantities. Operation-source summaries used the recorded action classes of retained reactions. Terminal-identity comparisons used canonical-SMILES sets within the same target. Representative routes were linked to their frozen records; the RF25 intervention example was selected by a deterministic post hoc ranking of the variation observed across conditions. The selection script and full case records accompany the analysis sources.

### S6.1 Statistical definitions

**Table S8 | Statistical definitions.**

| Parameter | Value |
|---|---|
| Bin edges | `[0, 0.2, 0.4, 0.6, 0.8, 1.000001]`; right-closed; lowest included |
| Empty target-bin | Undefined; excluded from that bin's mean |
| Resampling | 10,000 target bootstrap samples per cohort-bin; sample size equals number of contributing targets |
| Intervals | Pointwise 2.5th and 97.5th percentiles |
| Seed for the custom profile | `20260828 + dataset_index * 10 + bin_index`; PaRoutes120=0, RF25=1; bin_index=0..4 |
| Seed for the revised-plan profile | `20260928 + dataset_index * 10 + bin_index`; same cohort/bin indexing |

The target-level extraction records, figure-specific field mappings and bin-wise contributing-target counts are supplied in the machine-readable Source Data and analysis scripts. They are not reproduced as additional Supplementary Methods tables.

The illustrative RF25 target was selected by ranking targets using 100 times the number of distinct internal stopping types, plus 10 times the number of conditions with nonzero reaction counts, plus the number of distinct chemistry-source combinations, the maximum observed depth and the maximum unresolved-frontier count. Two points were added when native Rachel strictly closed the target. Ties were resolved by target ID. This deterministic score was used only to select an illustrative case; it was not a chemical-quality measure or an additional closure endpoint.

### S6.2 Route Atlas assembly

The Rachel Route Atlas connects the cohort-level outcomes to individual molecular routes in two offline HTML files, `PaRoutes120.html` and `RF25.html` (Supplementary Data; Figure S18). It includes the eight PaRoutes120 methods and four RF25 methods compared in the main text, with one position for every target-method combination. All 120 PaRoutes reference routes are included separately, using the adopted n1 and n5 subsets from Zenodo record 7341155. Reference routes are dataset comparators rather than additional planning runs and are not assigned a method strict-closure outcome. The 1,180 display positions comprise 1,060 method positions and 120 references; they are not counts of successful routes.

Route selection follows the frozen Fig. 3 endpoint tables and method-specific inventories. The selected final route or frozen process tree is displayed together with its recorded endpoint. Partial routes and positions without route output remain visible, and the adopted duplicate-generation exclusion for direct LLM target n5_7593 is retained. Native trees preserve molecular-node identities, nested trees preserve molecule occurrences, and flat reaction lists are connected by stereochemistry-aware canonical structure identity. Original reaction SMILES remain accessible; the identity-equivalent transformation recorded for direct LLM n5_4328 is retained as distinct source states.

Forward-model evidence is attached only when both the canonical product and the multiset of canonical reactants match the displayed reaction with stereochemistry retained. Whole-route appraisals require a match of the complete reaction multiset. Terminal audits are linked to their recorded molecular identities, and missing evidence remains unavailable. For Rachel targets n1_2664 and n5_2779, the frozen process routes contain 6 and 11 reactions, whereas the reviewed variants contain 5 and 10. These chemically distinct versions are displayed separately: the appraisal follows the evaluated variant, which does not automatically inherit the process tree's terminal audit or strict-closure label.

The companion preserves source hashes and reconciliation notes alongside the normalized records. Where current SyntheLite or AOT files differ from historical hashes, adopted chemistry is checked against frozen evaluator inputs or preserved partial-route views; chemical agreement is distinguished from byte identity. RF25 Rachel target rf25_012 (TMN-CPG) uses the adopted eight-reaction extract with the recorded original tree hash; the original session, tree and terminal files remain unavailable. The Atlas adds a way to inspect the adopted evidence and does not rerun planning or evaluation. Its JSON records, route index, version-difference ledger and checksum verifier accompany the HTML files; the release guide links method endpoints to the figure-level Source Data.

## Data and code availability

The Rachel Route Atlas is available on GitHub, with a fixed repository version 57c1716. It provides offline route comparisons for all 120 PaRoutes targets and 25 RF25 targets, including separate PaRoutes reference routes. Figure-level data, machine-readable records and supporting documentation are available in the Source Data release, version 2026-09-07. Rachel-authored materials are distributed under CC BY-NC 4.0; third-party terms are retained.

Rachel code, installation instructions and user guidance are available at https://github.com/ChazenLi/Rachel, with the present release identified by repository version 57c1716. The repository license is CC BY-NC 4.0; third-party dependencies retain their own terms. Experimental protocols and recorded configurations are described in Methods and Supporting Information.

## Figure S1

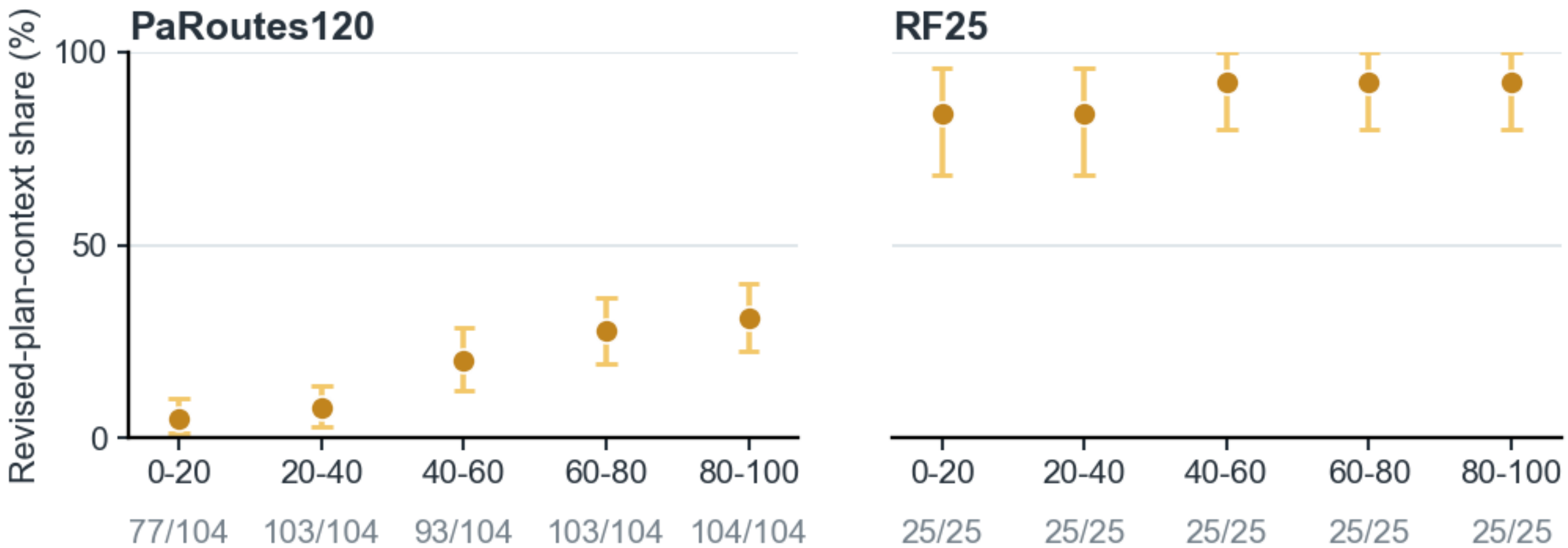

**Figure S1 | Revised-plan context across relative route position.** PaRoutes120 and RF25 are shown separately in five equal-width position bins. Each reaction's position is its order in the committed sequence divided by the route's total reaction count. Amber points show the equally weighted mean within-target fraction of committed steps registered under a revised plan; error bars show pointwise 95% intervals from 10,000 target-bootstrap resamples. Labels give contributing targets over routes with available plan histories: 104 PaRoutes120 routes and 25 RF25 routes. Targets without a step in a bin do not contribute to that bin's mean. This eligible population differs from the revised-route subset in Fig. 4c. Registered plan context does not establish a causal effect of revision.

## Figure S2

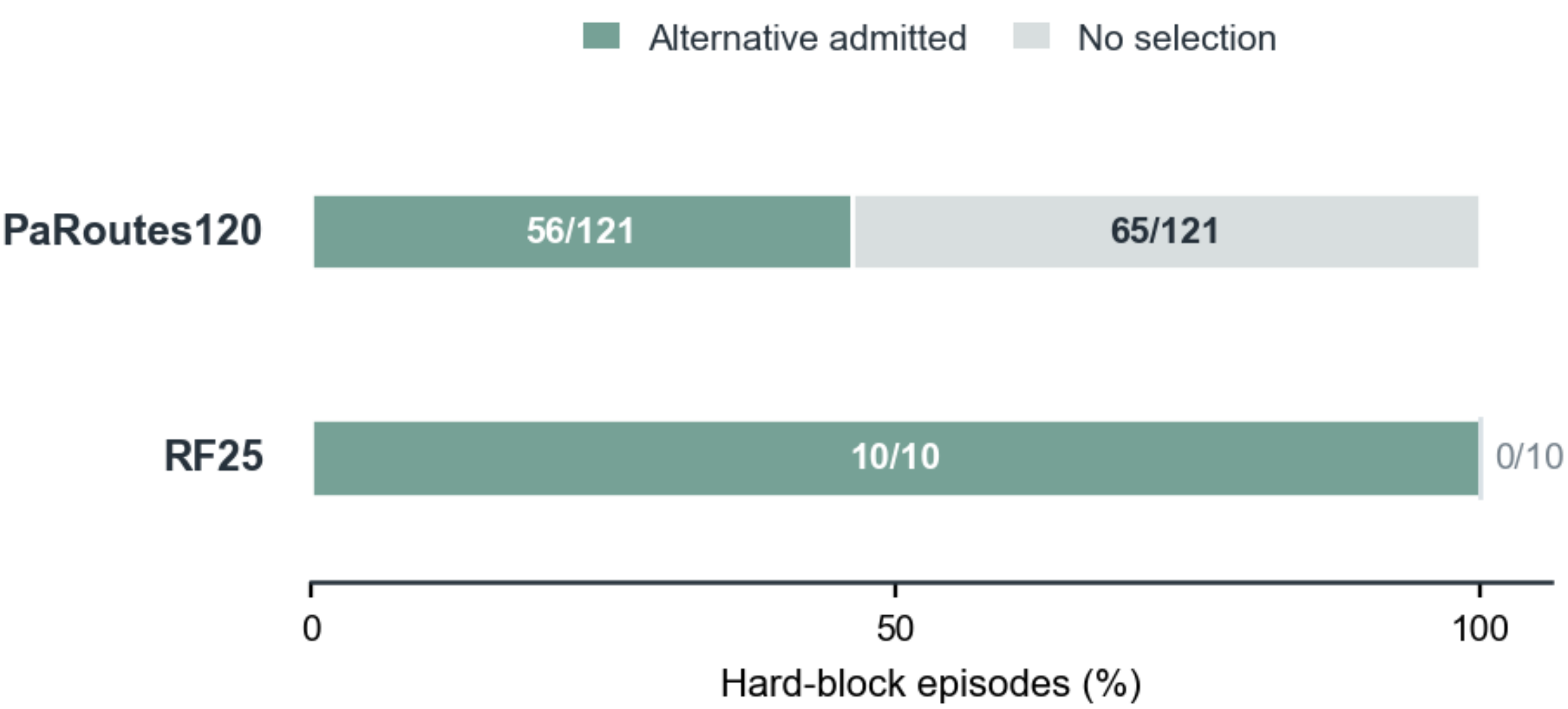

**Figure S2 | Decisions containing hard-blocked candidate records.** Decision episodes containing at least one hard-block record are divided into those in which another candidate was admitted to the final route and those with no selection in the episode. Bars show 56/121 and 65/121 episodes in PaRoutes120, and 10/10 and 0/10 in RF25, respectively. Admission refers to the alternative candidate, not the hard-blocked record. Episodes are nested within target routes; these counts describe recorded outcomes rather than independent replicates or causal gate efficacy.

# Figure S3

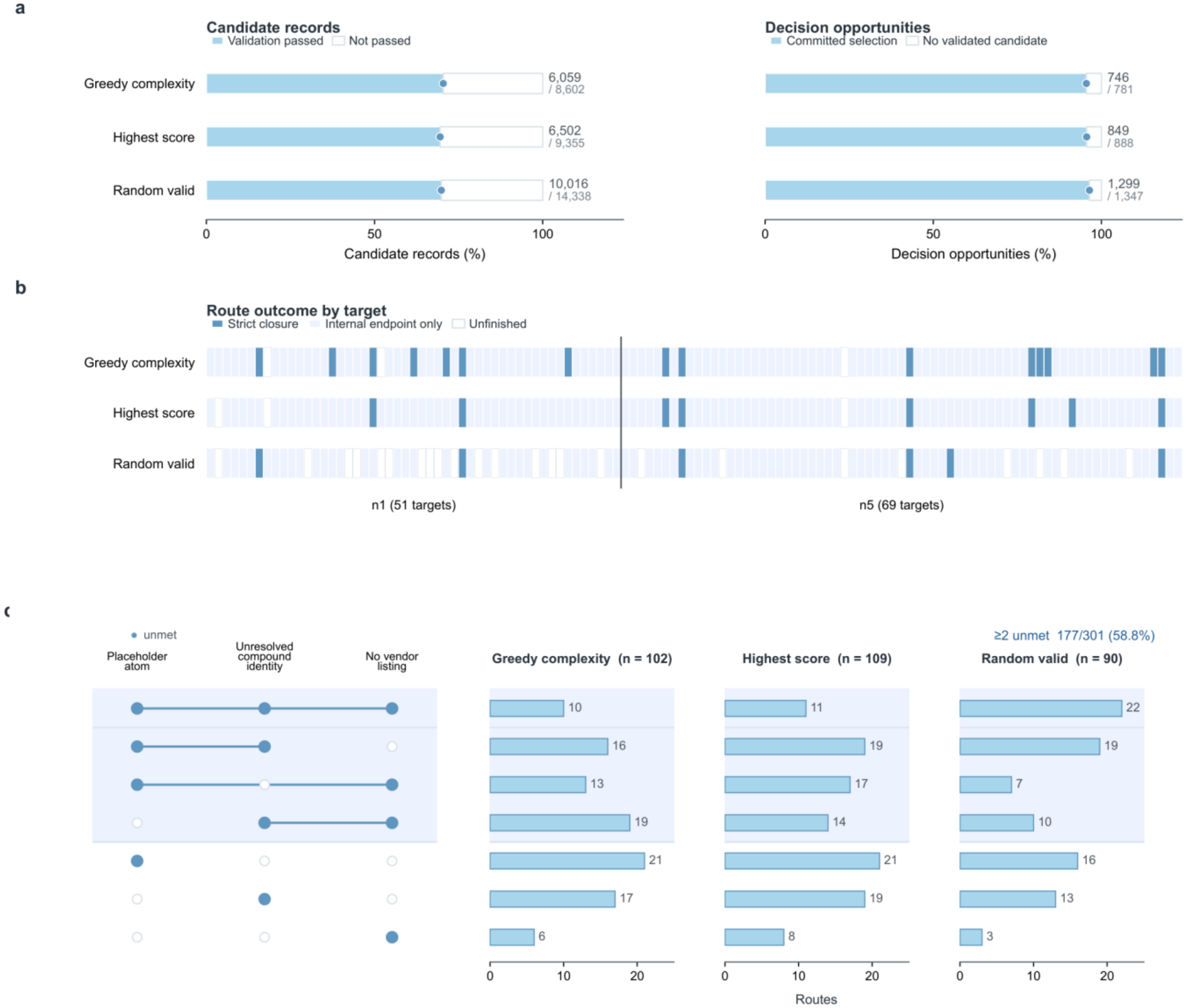

**Figure S3 | PaRoutes120 replacement audit. a,b,** Candidate records, candidate-bearing decision opportunities and policy-target route endpoints under the three mechanical replacements. **c,** Observed combinations of unmet placeholder-atom, unresolved-compound-identity and vendor-listing requirements among internally complete but non-strict mechanical routes. Candidate records, decision opportunities and target routes are distinct reporting units. Internal completion and detailed audit failures are diagnostic categories; the main route endpoint remains Strict Closure or Unresolved. The audits do not form a sequential funnel or establish mediation.

# Figure S4

Intervention realization

| | Restricted signal | | Retained control | | Execution check | |
|---|---|---|---|---|---|---|
| No custom action | Custom registration | 0/25 | Route plan | 25/25 | Completion required | 15/25 |
| No persistent plan | Route plan | 0/25 | Custom registration | 24/25 | Detailed validation | 25/25 |
| No sketch / continuation | Sketch / continuation | 0/25 | Route plan | 25/25 | Detailed validation | 25/25 |
| No detailed feedback | Detailed feedback | 0/25 | Private validation | 25/25 | Gate consistency | 25/25 |

Targets 01-25

**Figure S4 | Target-level realization of the four RF25 restrictions.** Each circle denotes one target. The left column records the restricted signal, the middle column a retained control and the right column an execution check. In row order, the three target counts are 0/25, 25/25 and 15/25 for No custom action; 0/25, 24/25 and 25/25 for No persistent plan; and 0/25, 25/25 and 25/25 for each of No sketch/continuation and No detailed feedback. The respective signals and checks are custom registration/route plan/completion-required events; route plan/custom registration/detailed validation; sketch or continuation/route plan/detailed validation; and model-visible detailed feedback/private validation/matched gate records. Here Restricted counts recorded use of the removed signal; the compact check in Fig. 5c instead displays its absence. The columns contain condition-specific variables, and observed use is distinct from command availability.

## Figure S5

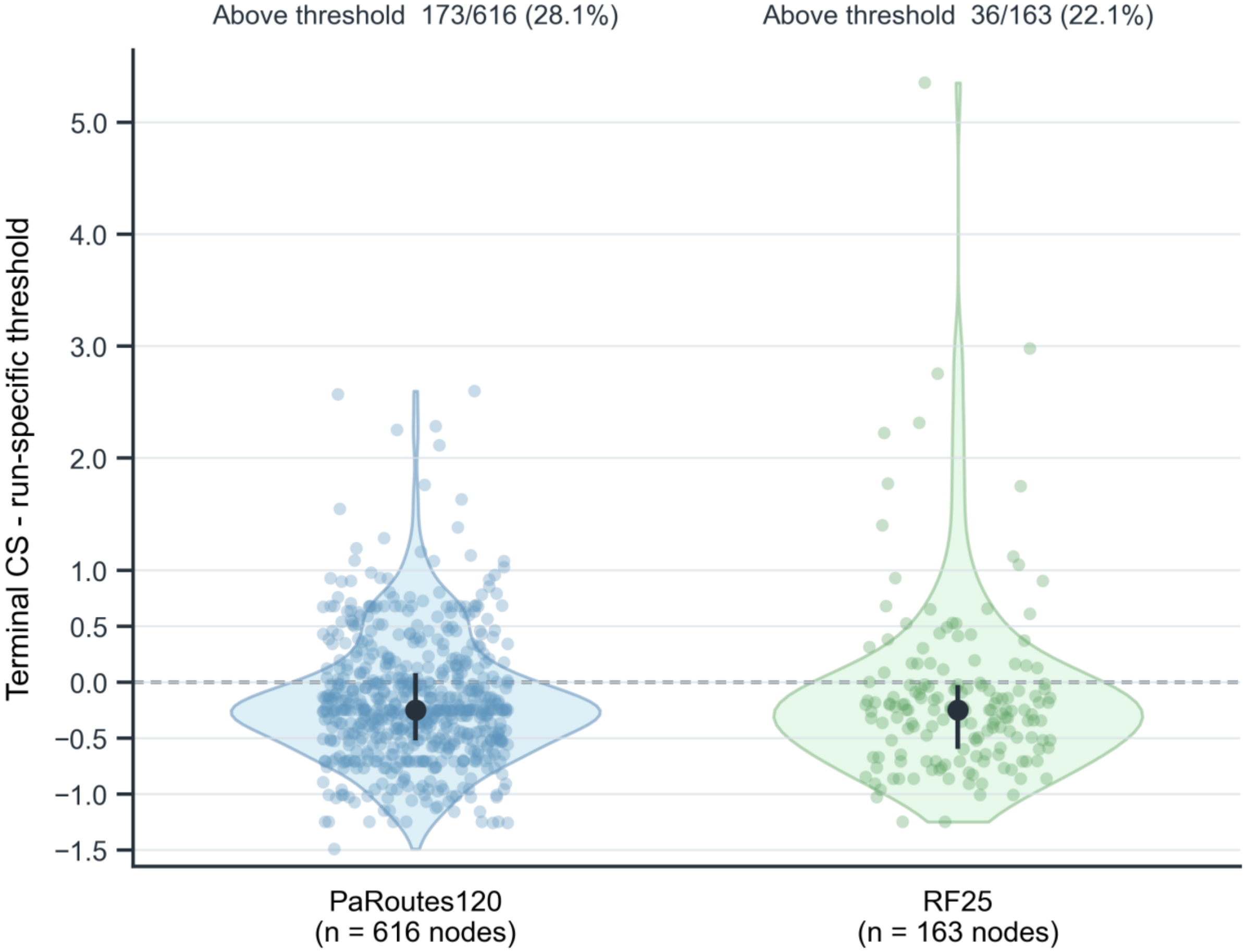

**Figure S5 | Final-terminal complexity relative to the planning threshold.** Distributions of terminal molecular complexity score (CS) minus the threshold configured for the corresponding run, shown separately for PaRoutes120 (616 terminal nodes) and RF25 (163). Points represent terminal nodes; the violin displays the distribution and neutral summaries indicate the median and interquartile range. Values exceed zero for 173/616 PaRoutes120 nodes and 36/163 RF25 nodes. Nodes are nested within routes. The planning threshold is distinct from the independent terminal-source criterion used to determine strict closure.

## Figure S6

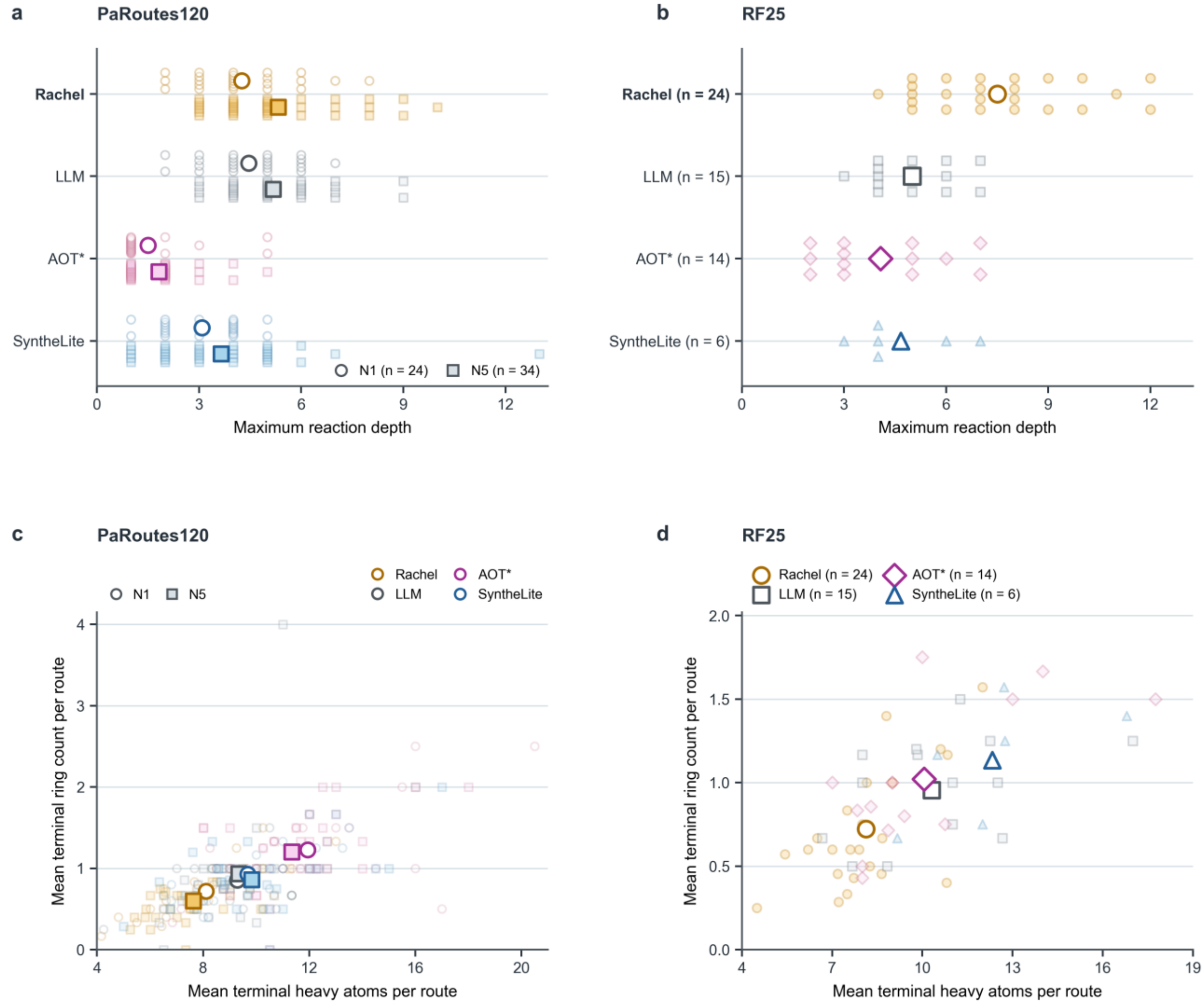

**Figure S6 | Route morphology conditional on strict closure.** Maximum reaction depth in PaRoutes120 (**a**) and RF25 (**b**), and mean terminal heavy-atom count versus mean terminal ring count per route in the same cohorts (**c**,**d**). PaRoutes120 contains four methods on independent n1 and n5 common strictly closed sets (24 and 34 targets per method; 232 method-route records). RF25 uses method-specific strictly closed sets: Rachel, 24; LLM, 15; AOT*, 14; SyntheLite, 6 routes. Pale marks represent individual routes and larger marks represent arithmetic means; colours identify methods, with n1/n5 distinguished by symbols. RF25 populations are unmatched. These are descriptive conditional views, not measures of synthetic feasibility or paired method effects.

# Figure S7

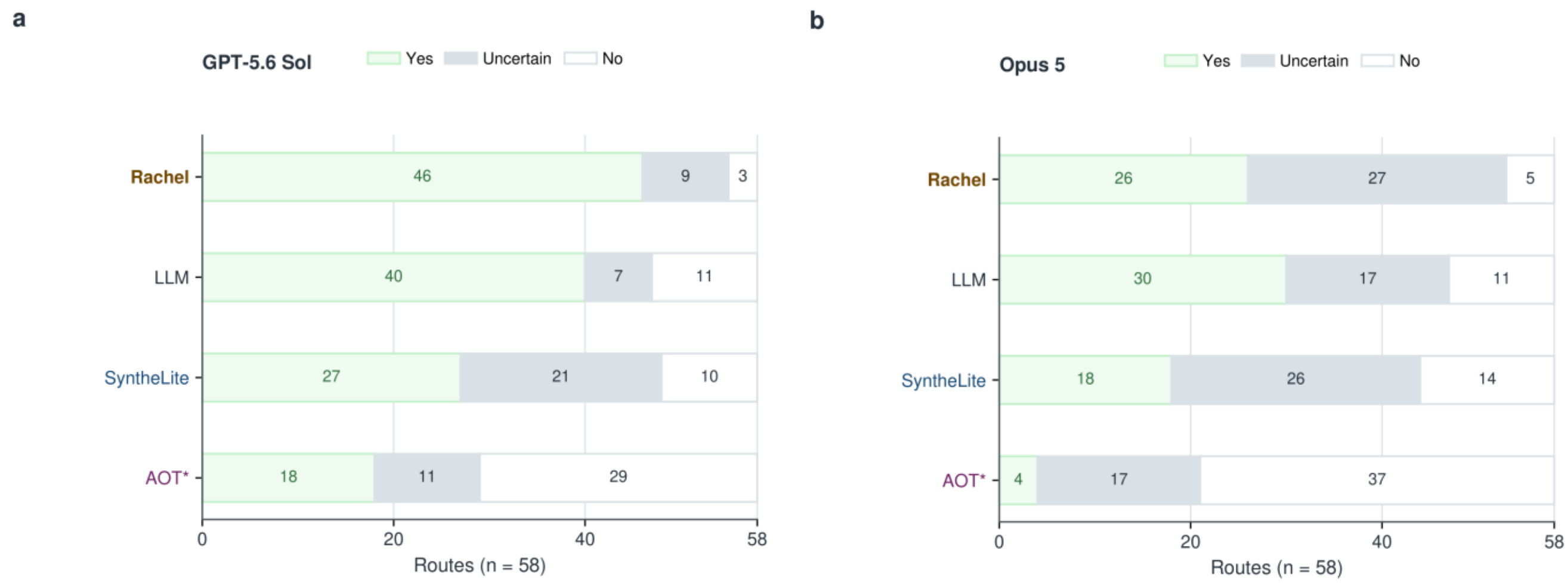

**Figure S7 | Reviewer-specific advancement decisions for common58 routes.** Complete yes, uncertain and no decision compositions are shown separately for GPT-5.6 Sol (**a**) and Opus 5 (**b**) across four route-construction methods. Each method row contains 58 method-blinded whole-route appraisals. The evaluators were analysed separately and were not pooled. These computational recommendations concern progression to human synthetic review and do not constitute that review or establish experimental feasibility.

## Figure S8

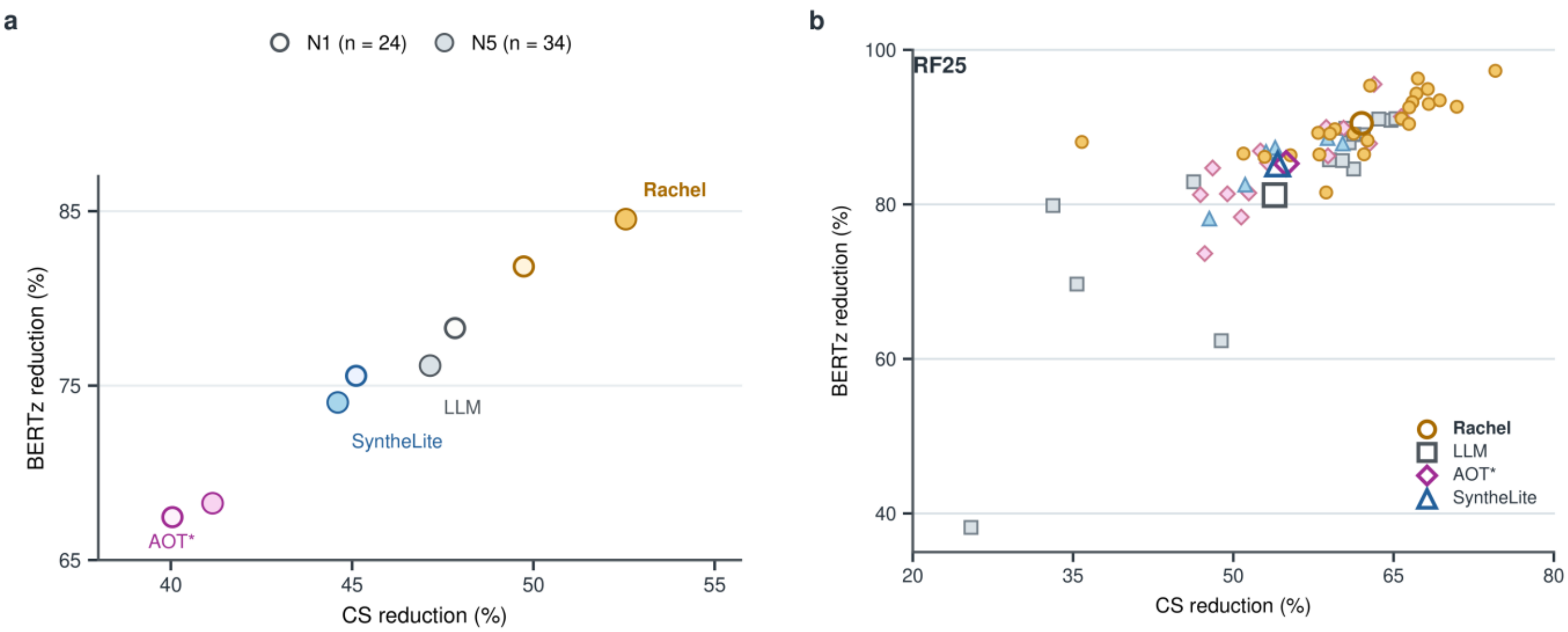

**Figure S8 | CS and BERTz reduction profiles of strictly closed routes. a,** Mean target-to-terminal CS reduction versus mean target-to-terminal BERTz reduction for four methods in the independent PaRoutes120 n1 (24 targets per method) and n5 (34 targets per method) common strictly closed cohorts. Marker fill denotes cohort and colour denotes method; no connector is drawn because n1 and n5 are not paired. **b,** Target-to-terminal CS and BERTz percentage reductions for strictly closed RF25 routes. Pale symbols show individual routes and larger hollow symbols show method means (Rachel, n = 24; LLM, 15; AOT*, 14; SyntheLite, 6); route sets are unmatched. Both panels describe route simplification rather than experimental feasibility.

**Figure S9**

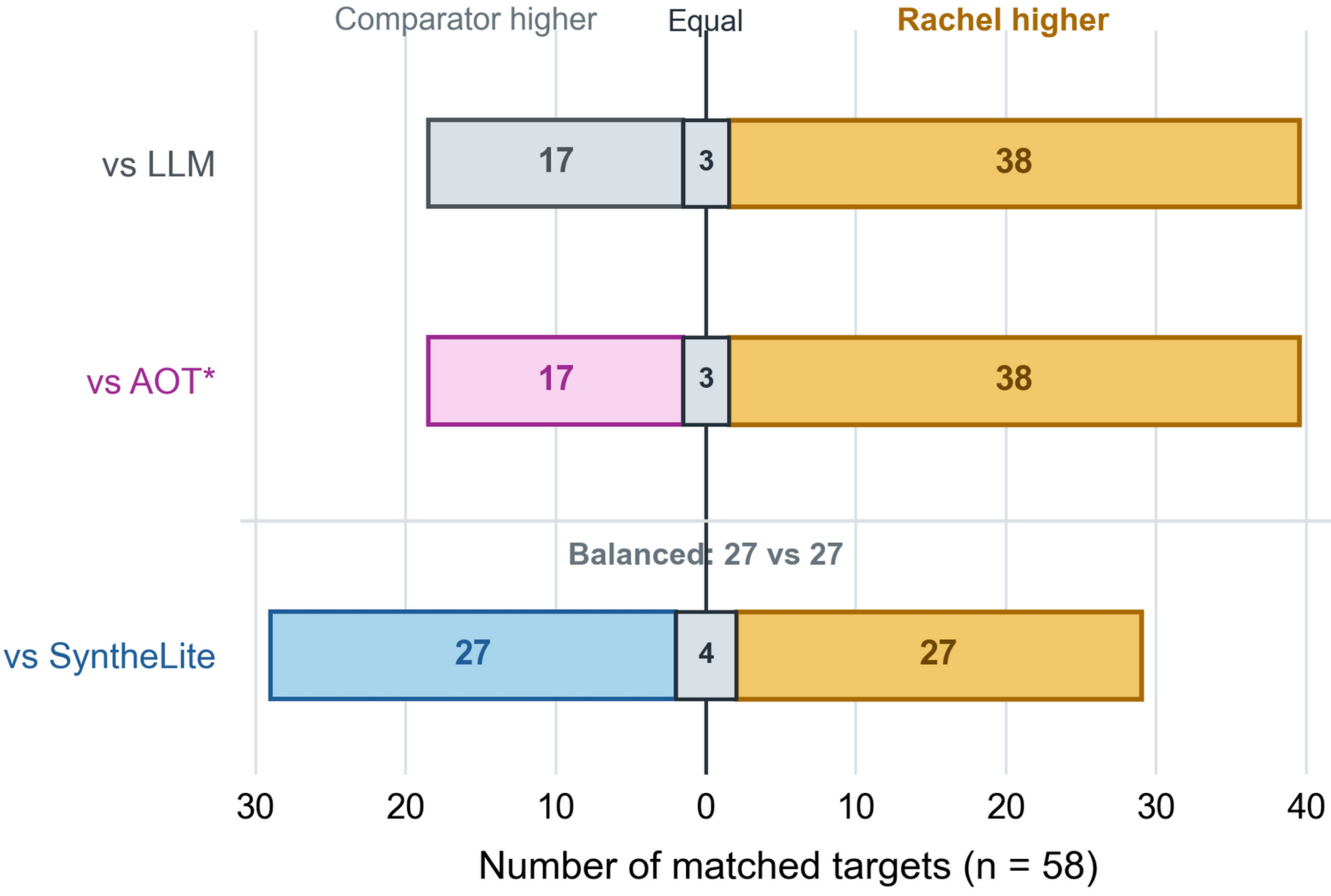

**Figure S9 | Target-level Both@10 comparisons on common58.** Counts of matched common58 targets on which Rachel's route-level Both@10 is lower than, equal to or higher than each comparator (n = 58 per comparison). Bars encode target counts, not difference magnitudes.

## Figure S10

Panel a

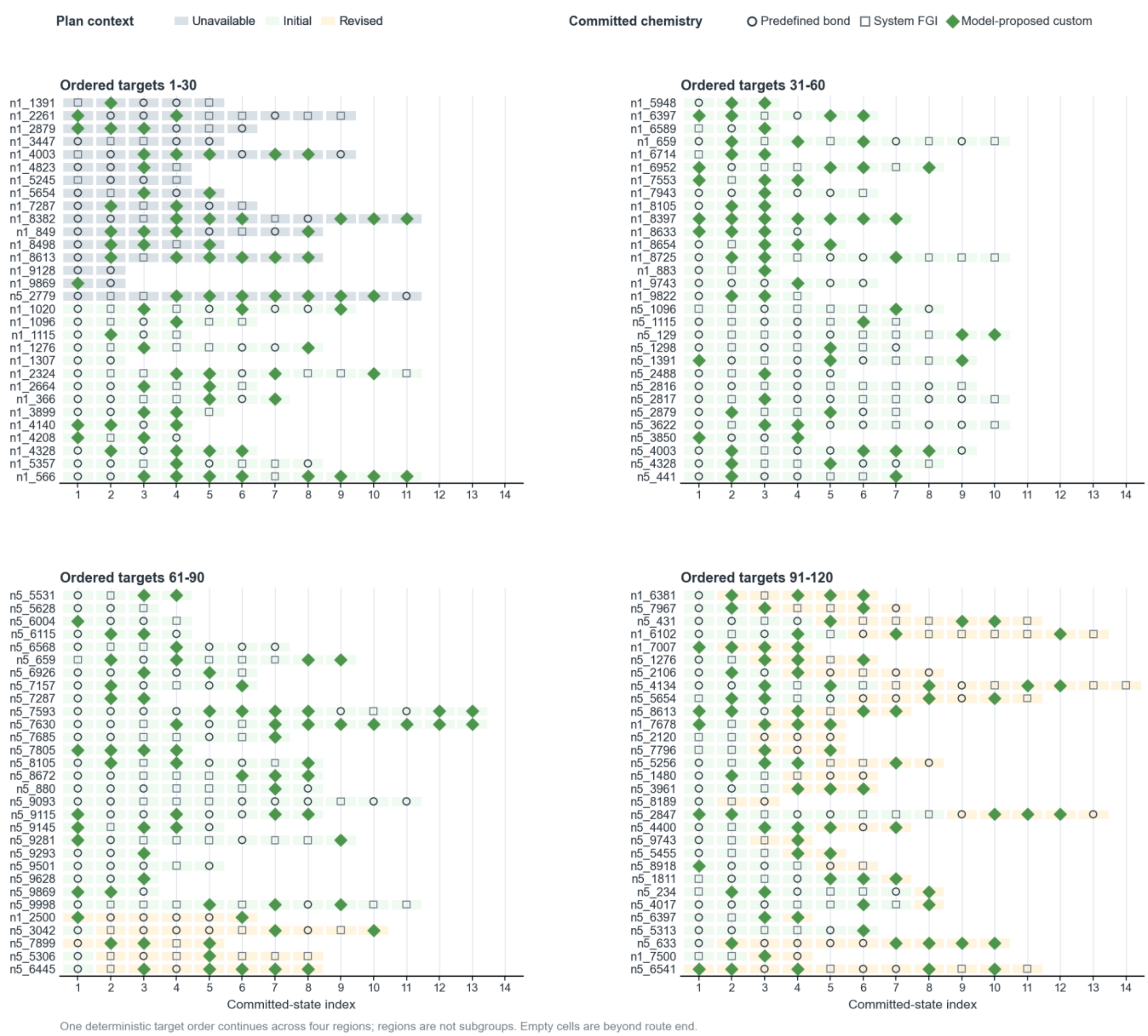

**Figure S10 | Complete route-level records of committed chemistry. a,** PaRoutes120, 806 committed transitions across 120 targets. **b,** RF25, 325 transitions across 25 targets. Each row represents one route and horizontal position denotes actual committed-state index. Circles denote predefined bond operations, squares predefined functional-group interconversions and green diamonds model-proposed custom chemistry. Cell backgrounds indicate unavailable, initial or revised plan context. Empty space beyond a route's last step is route-tail space. The four PaRoutes120 blocks are pagination, not subgroups. The 1,131 transitions are nested within 145 routes; registered order is not elapsed time.

## Figure S10 (continued)

Panel b

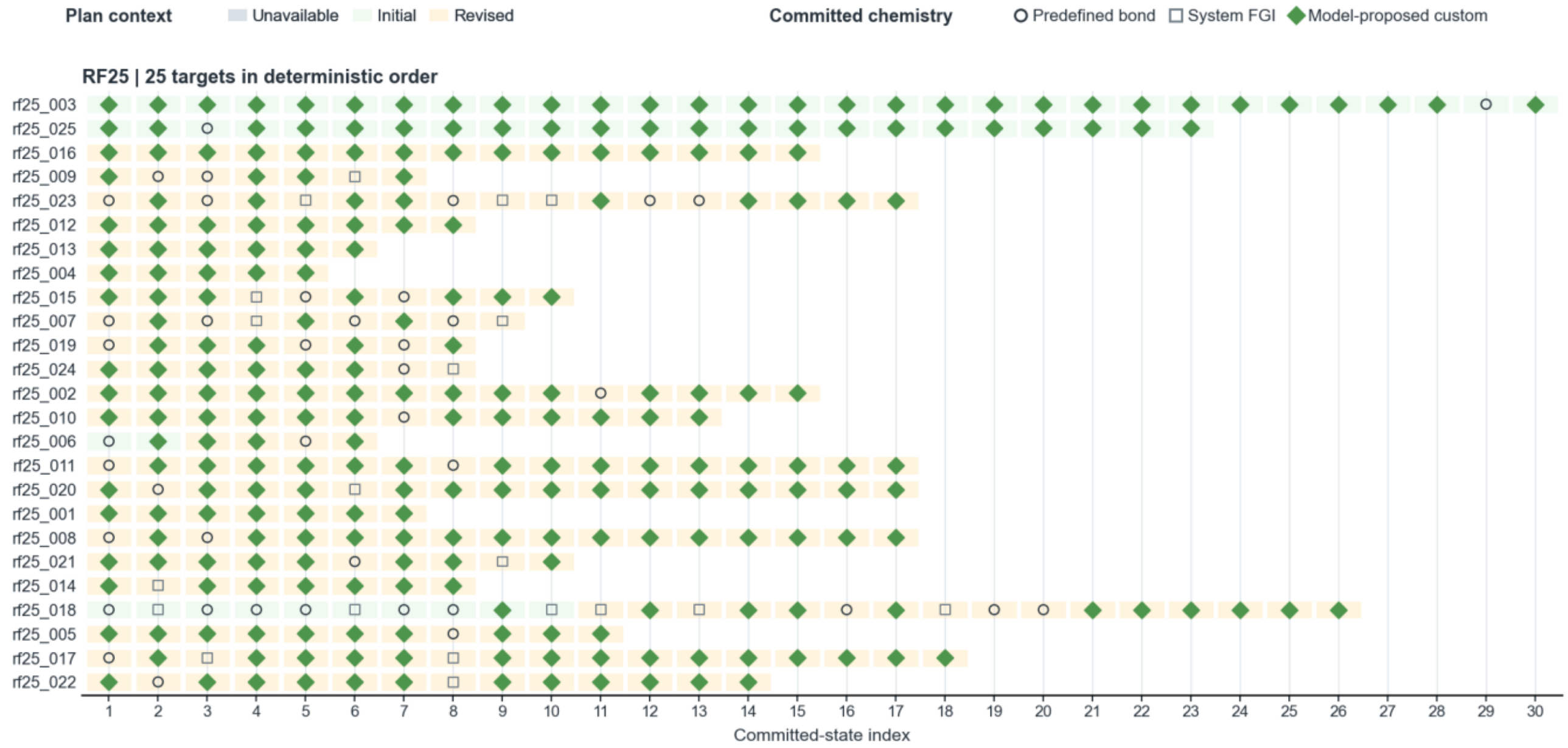

**Figure S10 | Complete route-level records of committed chemistry. a,** PaRoutes120, 806 committed transitions across 120 targets. **b,** RF25, 325 transitions across 25 targets. Each row represents one route and horizontal position denotes actual committed-state index. Circles denote predefined bond operations, squares predefined functional-group interconversions and green diamonds model-proposed custom chemistry. Cell backgrounds indicate unavailable, initial or revised plan context. Empty space beyond a route's last step is route-tail space. The four PaRoutes120 blocks are pagination, not subgroups. The 1,131 transitions are nested within 145 routes; registered order is not elapsed time.

## Figure S11

Panel a

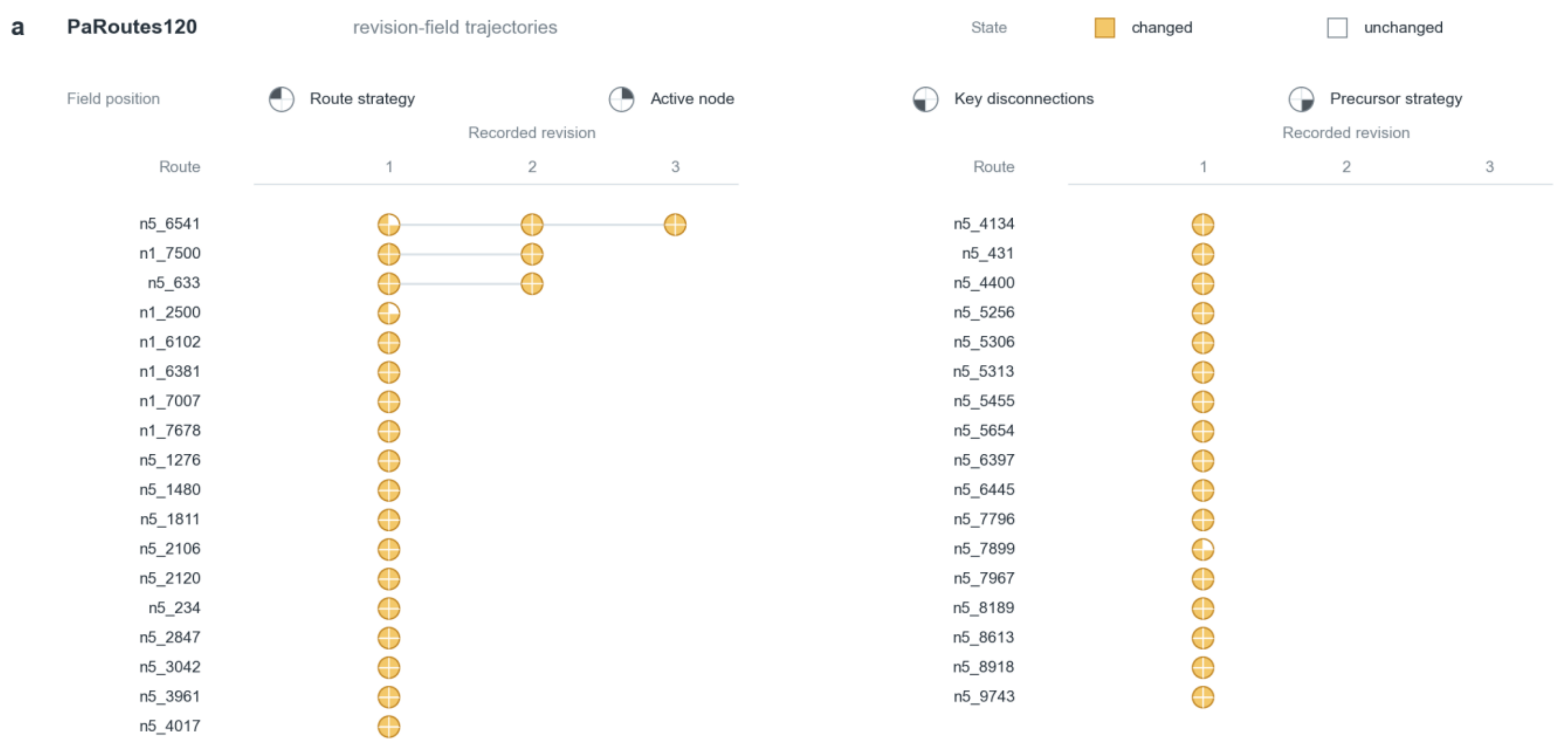

**Figure S11 | Route-by-event trajectories of recorded plan revisions. a,** PaRoutes120 revised routes. **b,** RF25 revised routes. Each row is one revised route, with revision events ordered horizontally. Each event has four quadrants identifying the registered active-node, key-disconnection, precursor-logic and route-thesis fields. Amber denotes a literal field change and white denotes no displayed change. The figure contains 58 revised routes, 145 revision events and 580 field comparisons. Three RF25 events with unresolved sources also appear white; their source_unresolved status is retained in Source Data, so white does not necessarily indicate a verified absence of change. Field turnover does not measure semantic distance or chemical importance.

**Figure S11 (continued)**

Panel b

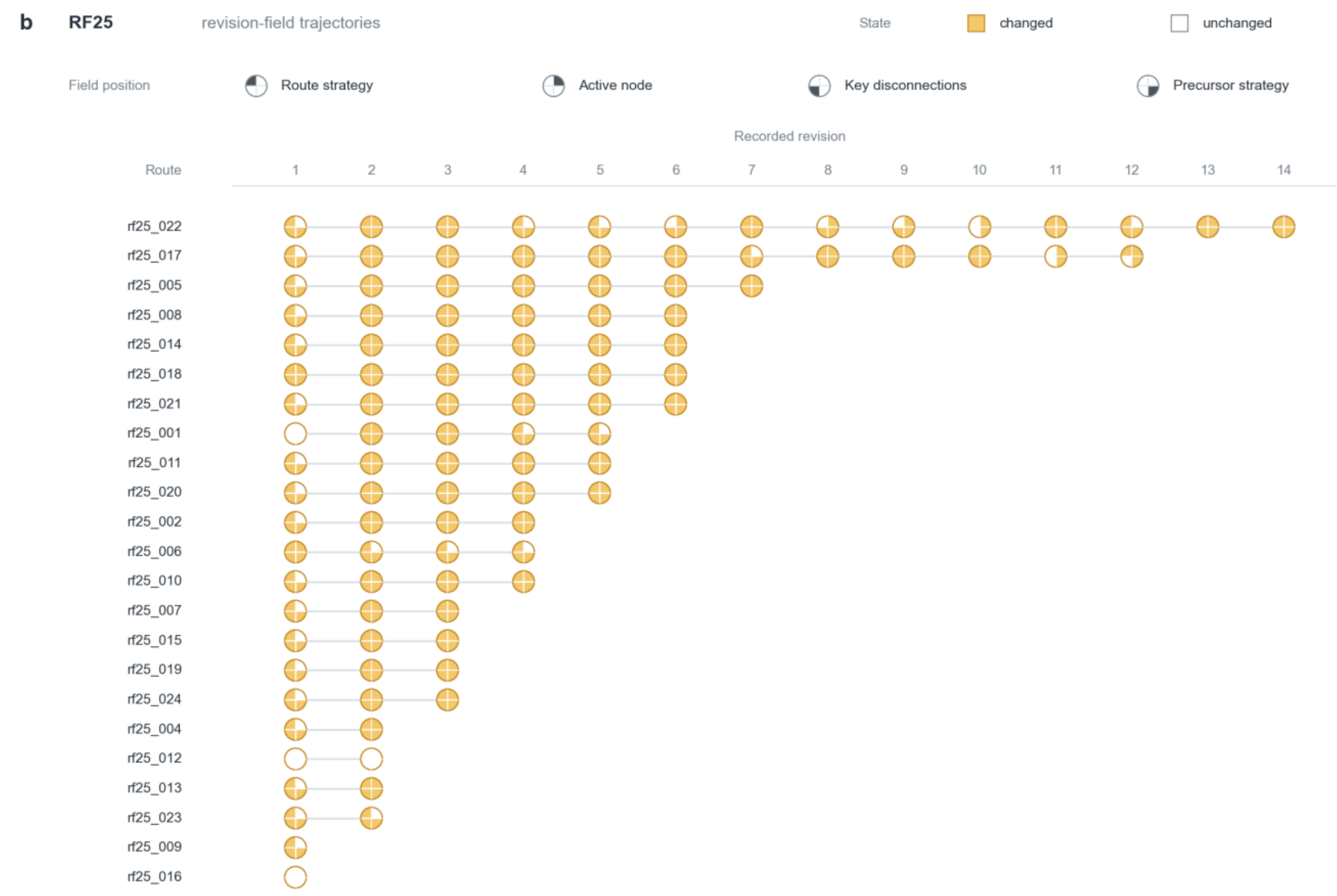

**Figure S11 | Route-by-event trajectories of recorded plan revisions. a,** PaRoutes120 revised routes. **b,** RF25 revised routes. Each row is one revised route, with revision events ordered horizontally. Each event has four quadrants identifying the registered active-node, key-disconnection, precursor-logic and route-thesis fields. Amber denotes a literal field change and white denotes no displayed change. The figure contains 58 revised routes, 145 revision events and 580 field comparisons. Three RF25 events with unresolved sources also appear white; their source_unresolved status is retained in Source Data, so white does not necessarily indicate a verified absence of change. Field turnover does not measure semantic distance or chemical importance.

## Figure S12

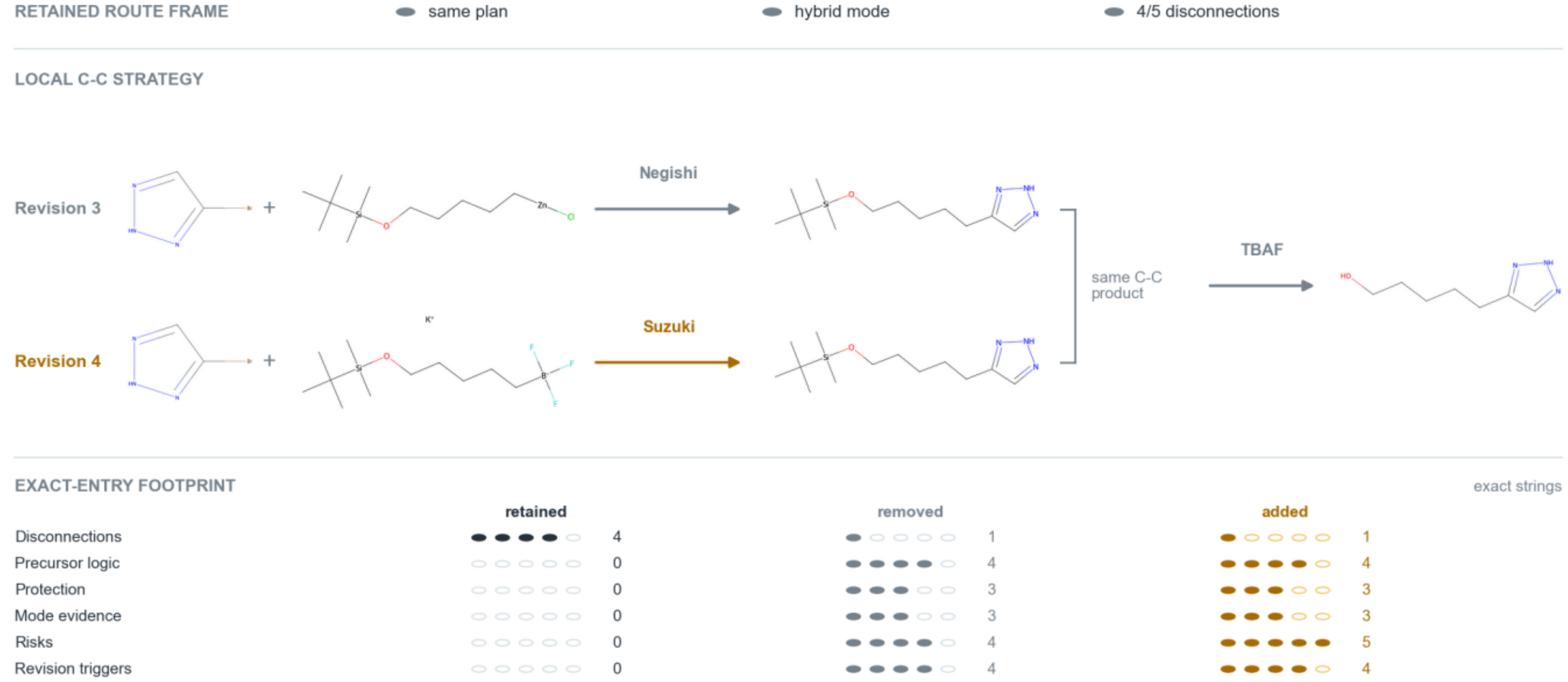

**Figure S12 | Chemical and field-level content of an independent plan revision.** RF25 route rf25_008 is compared between revisions 3 and 4. The recorded local strategy for the hydroxypentyl-azole carbon-carbon bond changes from Negishi/ZnCl to Suzuki/BF3K while the global route frame is retained. The accompanying ledger counts exactly retained, removed and added entries in the registered plan fields. The chemical schemes depict plan content; the revision itself is not a replacement of one molecular node by another, and exact-entry turnover does not measure semantic or chemical improvement.

# Figure S13

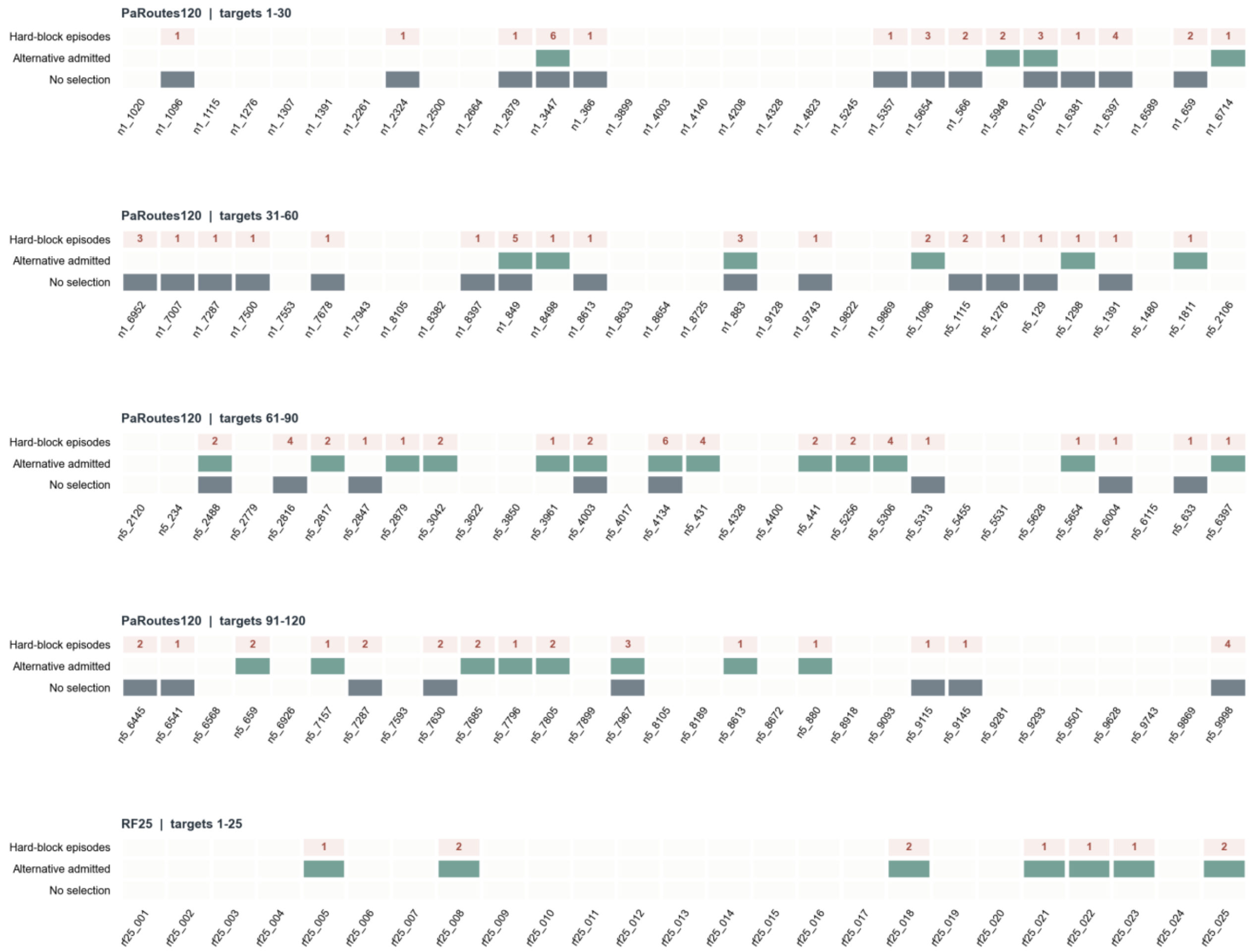

**Figure S13 | Complete target-level audit of hard-block-containing decisions.** All 145 registered targets are shown: PaRoutes120 in four 30-target strips and RF25 in one 25-target strip. Each column is one target. The first row gives the number of decision episodes containing a hard-block record. Filled cells in the second and third rows indicate whether at least one such episode admitted another candidate to the final route or had no selection, respectively. These two target-level indicators can both be present because a target can contain different episode outcomes. Figure S2 counts the episodes themselves. These records do not report target-level strict closure.

## Figure S14

Panel a

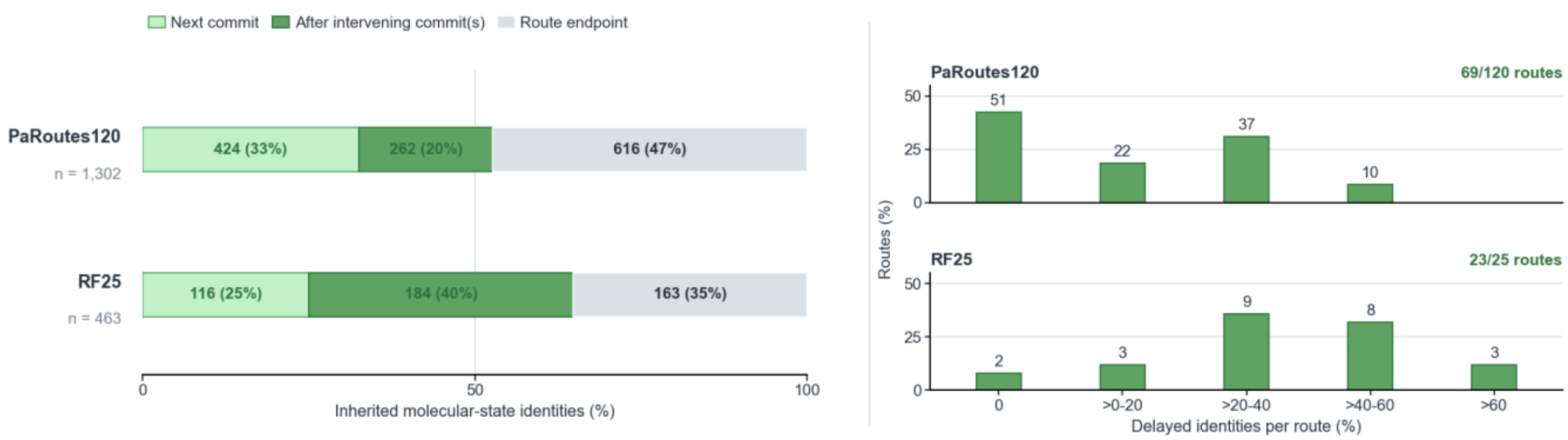

**Figure S14 | Outcomes of inherited molecular-state identities. a,** Identity-level outcome composition and route-level distribution of delayed-identity fractions for PaRoutes120 and RF25. Each unique inherited molecular-node identity is classified as active at the next commit, active after one or more intervening commits, or a final route endpoint. The respective counts are 424, 262 and 616 of 1,302 identities in PaRoutes120, and 116, 184 and 163 of 463 in RF25. **b,** Complete audit of all 145 routes, with one horizontal row per route showing the fractions of these three identity classes; the diamond to the right gives the total identity count. Molecular identities and routes are distinct reporting units. Registered intervals between commits do not measure elapsed time or establish learned memory or causal scheduling.

## Figure S14 (continued)

Panel b

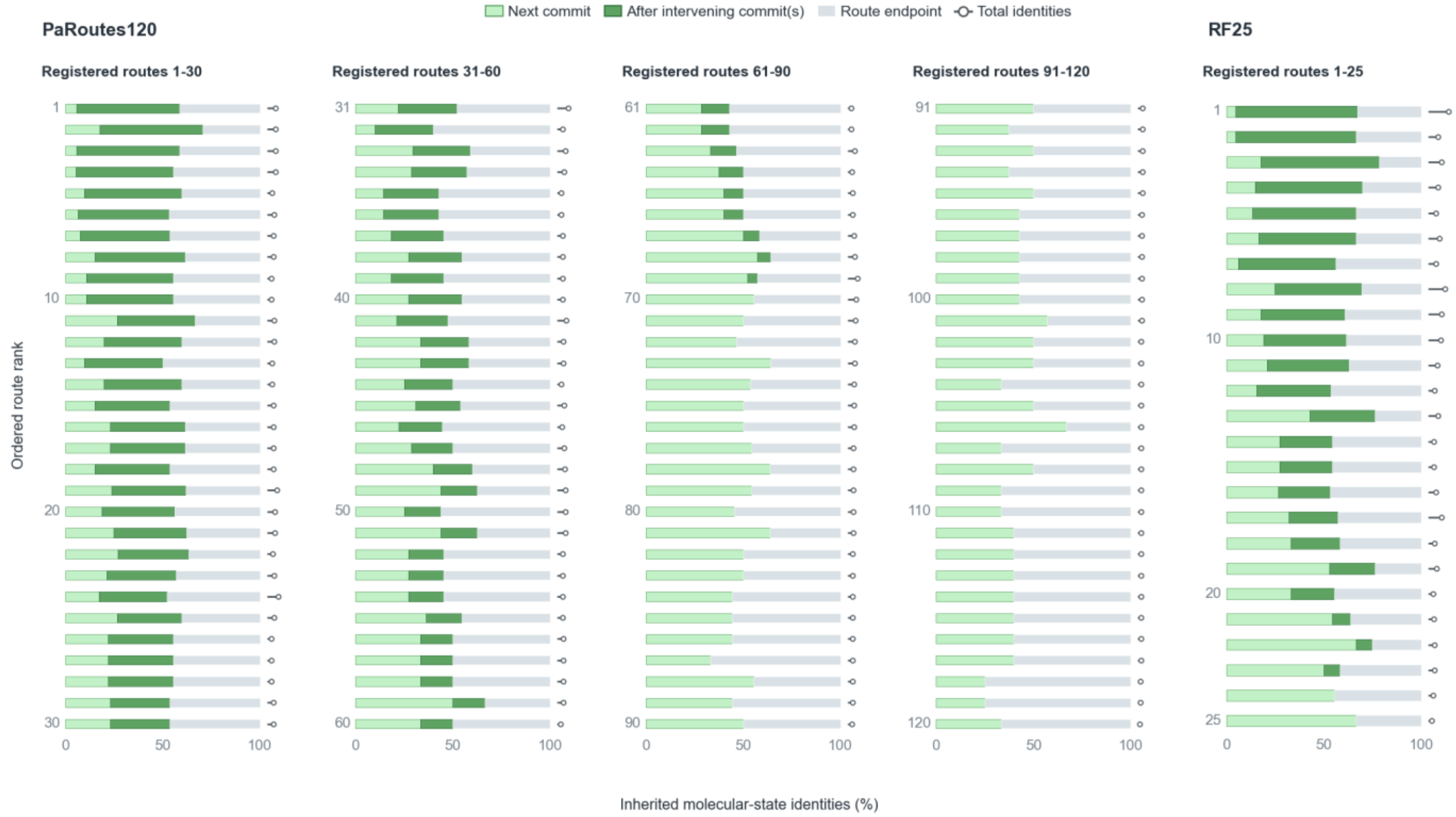

**Figure S14 | Outcomes of inherited molecular-state identities. a,** Identity-level outcome composition and route-level distribution of delayed-identity fractions for PaRoutes120 and RF25. Each unique inherited molecular-node identity is classified as active at the next commit, active after one or more intervening commits, or a final route endpoint. The respective counts are 424, 262 and 616 of 1,302 identities in PaRoutes120, and 116, 184 and 163 of 463 in RF25. **b,** Complete audit of all 145 routes, with one horizontal row per route showing the fractions of these three identity classes; the diamond to the right gives the total identity count. Molecular identities and routes are distinct reporting units. Registered intervals between commits do not measure elapsed time or establish learned memory or causal scheduling.

## Figure S15

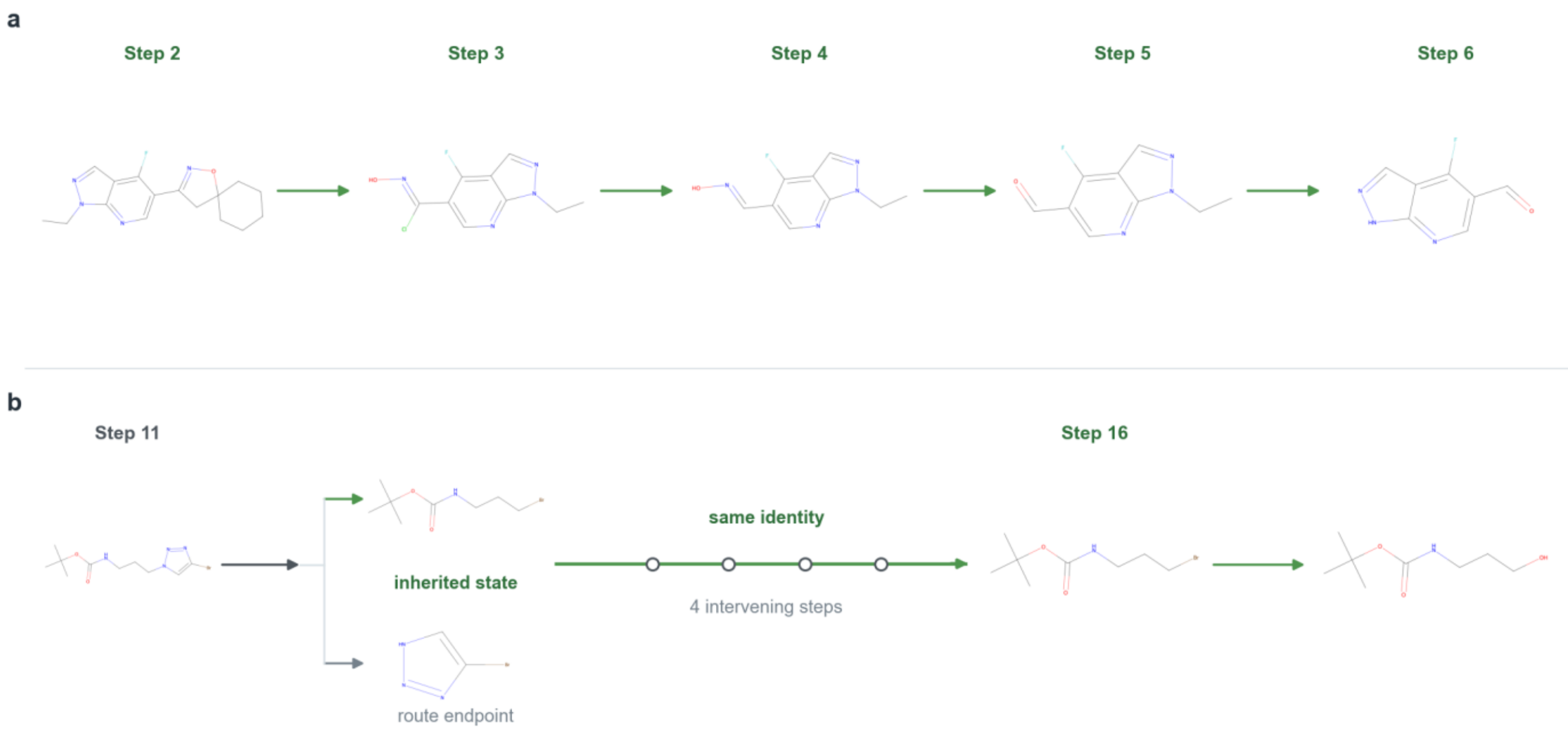

**Figure S15 | Consecutive and delayed inheritance of molecular-state identities. a,** Consecutive state sequence across Steps 2-6 of PaRoutes120 route n1_849, showing four successive identity handoffs. **b,** An identity written at Step 11 of RF25 route rf25_011 becomes the active product at Step 16 after four intervening registered steps; the sibling terminal is retained as a reference. The later step records Appel bromination of N-Boc-3-aminopropan-1-ol. Steps denote registered route order. The displayed structures follow the recorded candidate chemistry; full reaction records, including omitted co-reactants, are retained in Source Data. The cases illustrate identity linkage rather than experimental validation or population prevalence.

## Figure S16

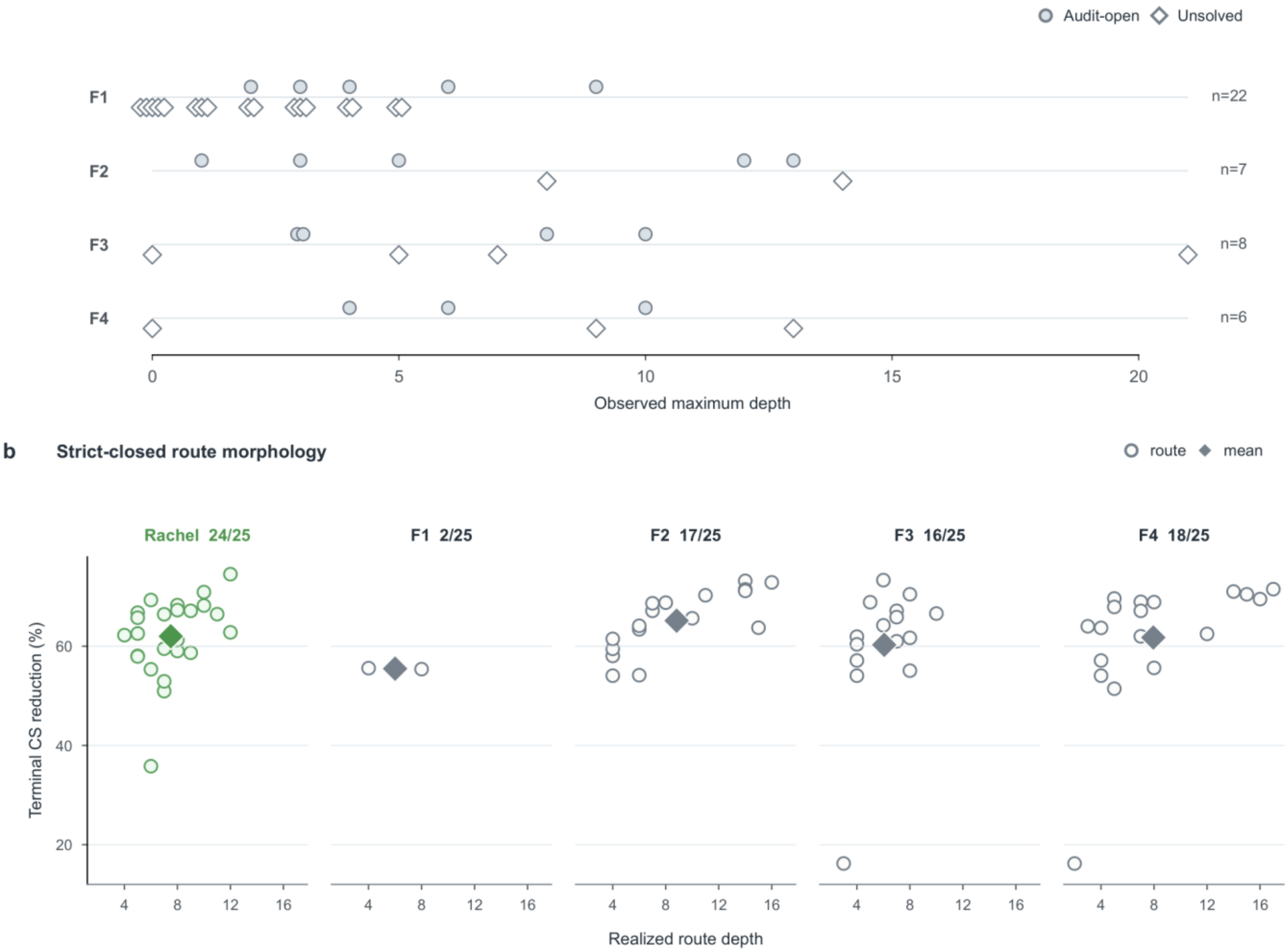

**Figure S16 | RF25 route morphology under functional restrictions.** F1-F4 denote No custom action, No persistent plan, No sketch/continuation and No detailed feedback, respectively; Skill Rachel denotes the native Rachel reference. **a,** Maximum observed depth of restricted routes that did not strictly close, within the 24 targets strictly closed by native Rachel. The four restrictions contribute 22, 7, 8 and 6 routes. Circles denote internally completed but audit-open routes and diamonds other unresolved routes. **b,** Depth and terminal-complexity reduction for the condition-specific strictly closed sets: native Rachel, 24 routes; F1-F4, 2, 17, 16 and 18 routes. Circles denote routes and diamonds condition means. The panels use outcome-conditioned populations; their points are not paired comparisons across conditions. Audit-open is a diagnostic subgroup of the main Unresolved endpoint. Depth and terminal complexity describe route morphology rather than efficiency or a restriction ranking.

## Figure S17

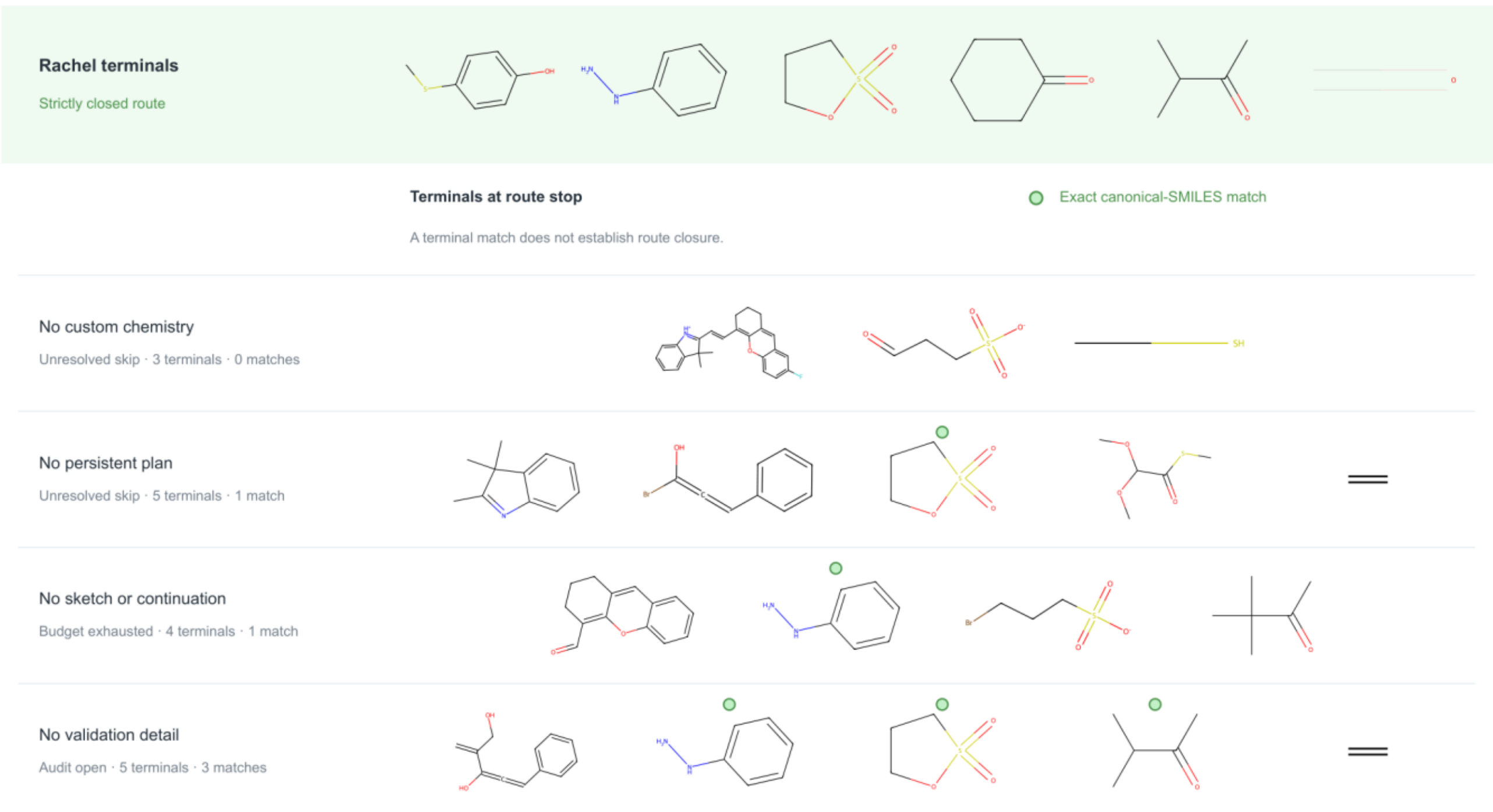

**Figure S17 | Terminal identities at route stop for target 014.** Rachel's six strictly closed terminal identities are shown above the terminal structures recorded when each restricted route stopped. Green marks identify exact canonical-SMILES matches to Rachel terminals. The four restricted terminal sets contain 3, 5, 4 and 5 leaves, of which 0, 1, 1 and 3 match, respectively. All four restricted routes remain Unresolved. A terminal match does not establish closure of the complete route; the structures come from one deterministically selected illustrative target.

# Figure S18

Panel a

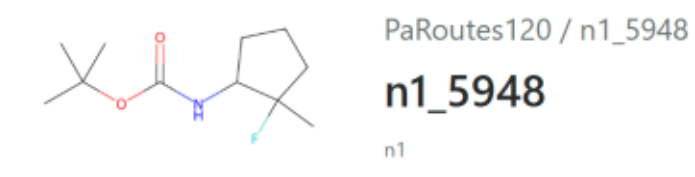

PaRoutes120 / n1_5948

**n1_5948**

n1

**Rachel**

• Strictly closed · 3 steps

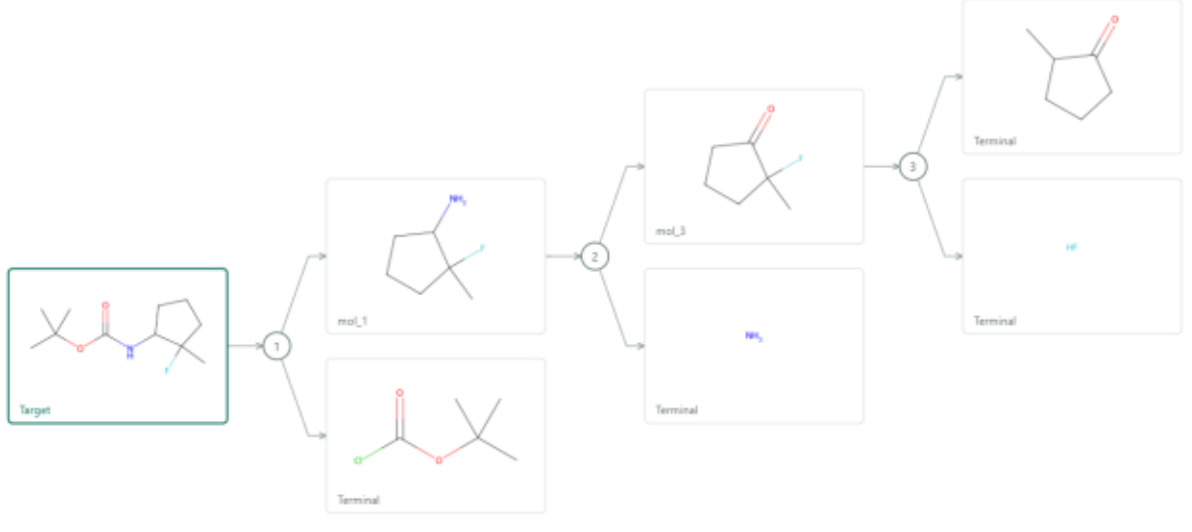

**Reference**

• PaRoutes reference route · 2 steps

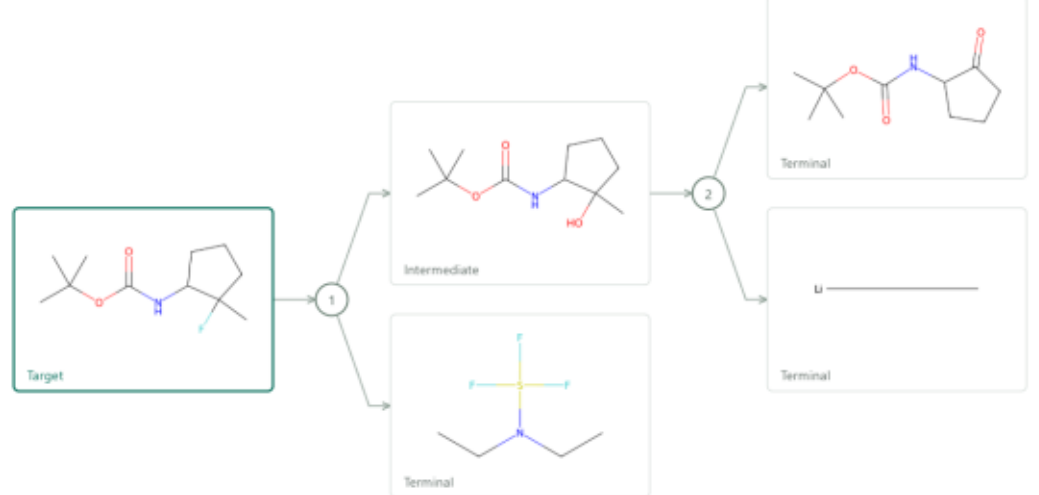

**Figure S18 | Individual routes behind the cohort comparisons. a,** Rachel and the PaRoutes reference route for target n1_5948. **b,** Rachel and direct LLM routes for RF25 target rf25_004 (D-BDY). These short route pairs were chosen for legibility, not as a representative sample. The views use the frozen offline Atlas, with method panels stacked and navigation controls omitted for print. Arrows follow retrosynthetic expansion from target to precursors; numbered circles identify recorded reactions. Closure labels reproduce the adopted audit outcomes; the Reference entry has no method-closure assignment. The complete Atlas includes every target-method position, including incomplete and unavailable outcomes, and permits inspection of molecular structures and matched evidence at full size (Supplementary Data; S6.2).

## Figure S18 (continued)

Panel b

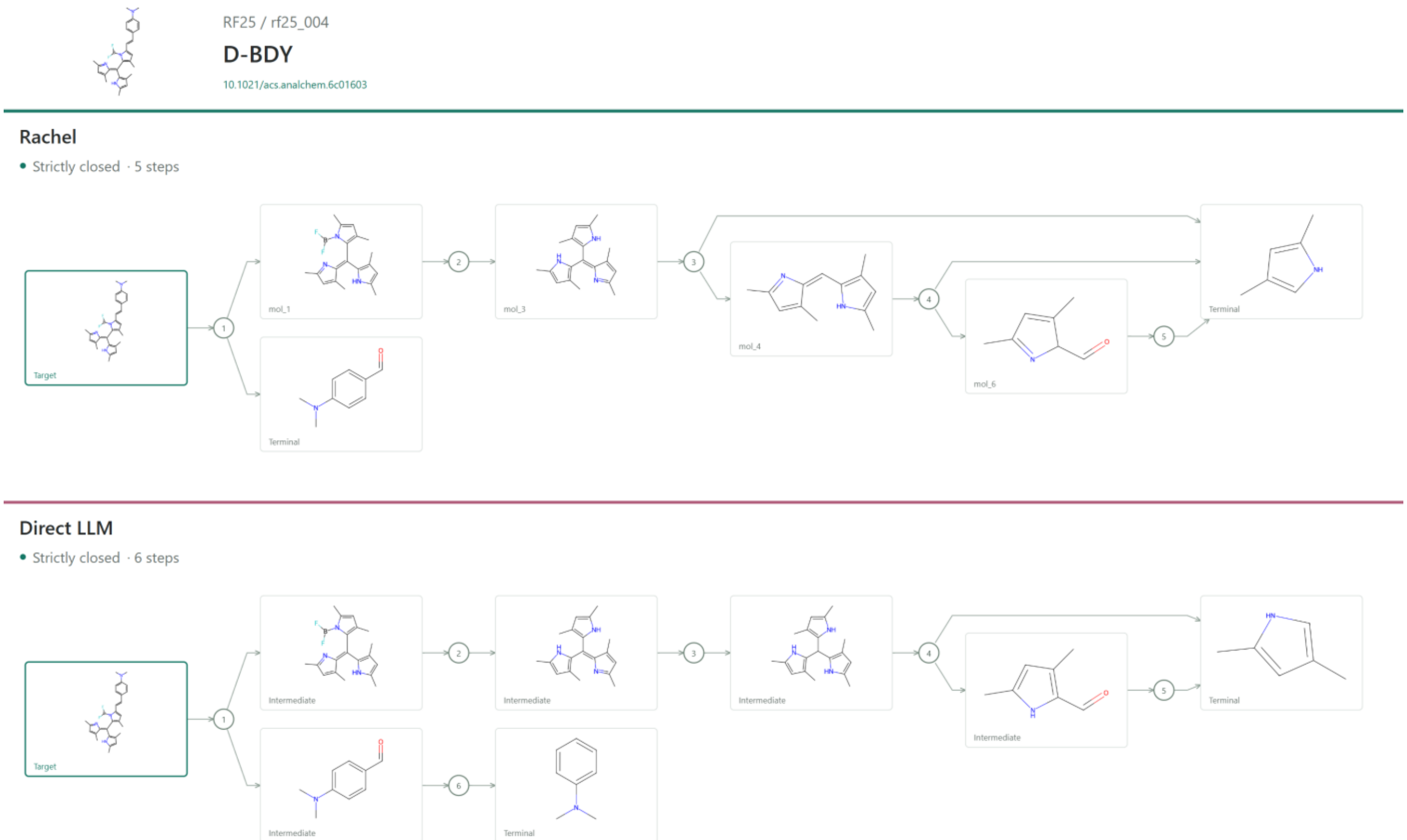

**Figure S18 | Individual routes behind the cohort comparisons. a,** Rachel and the PaRoutes reference route for target n1_5948. **b,** Rachel and direct LLM routes for RF25 target rf25_004 (D-BDY). These short route pairs were chosen for legibility, not as a representative sample. The views use the frozen offline Atlas, with method panels stacked and navigation controls omitted for print. Arrows follow retrosynthetic expansion from target to precursors; numbered circles identify recorded reactions. Closure labels reproduce the adopted audit outcomes; the Reference entry has no method-closure assignment. The complete Atlas includes every target-method position, including incomplete and unavailable outcomes, and permits inspection of molecular structures and matched evidence at full size (Supplementary Data; S6.2).

## Figure S19

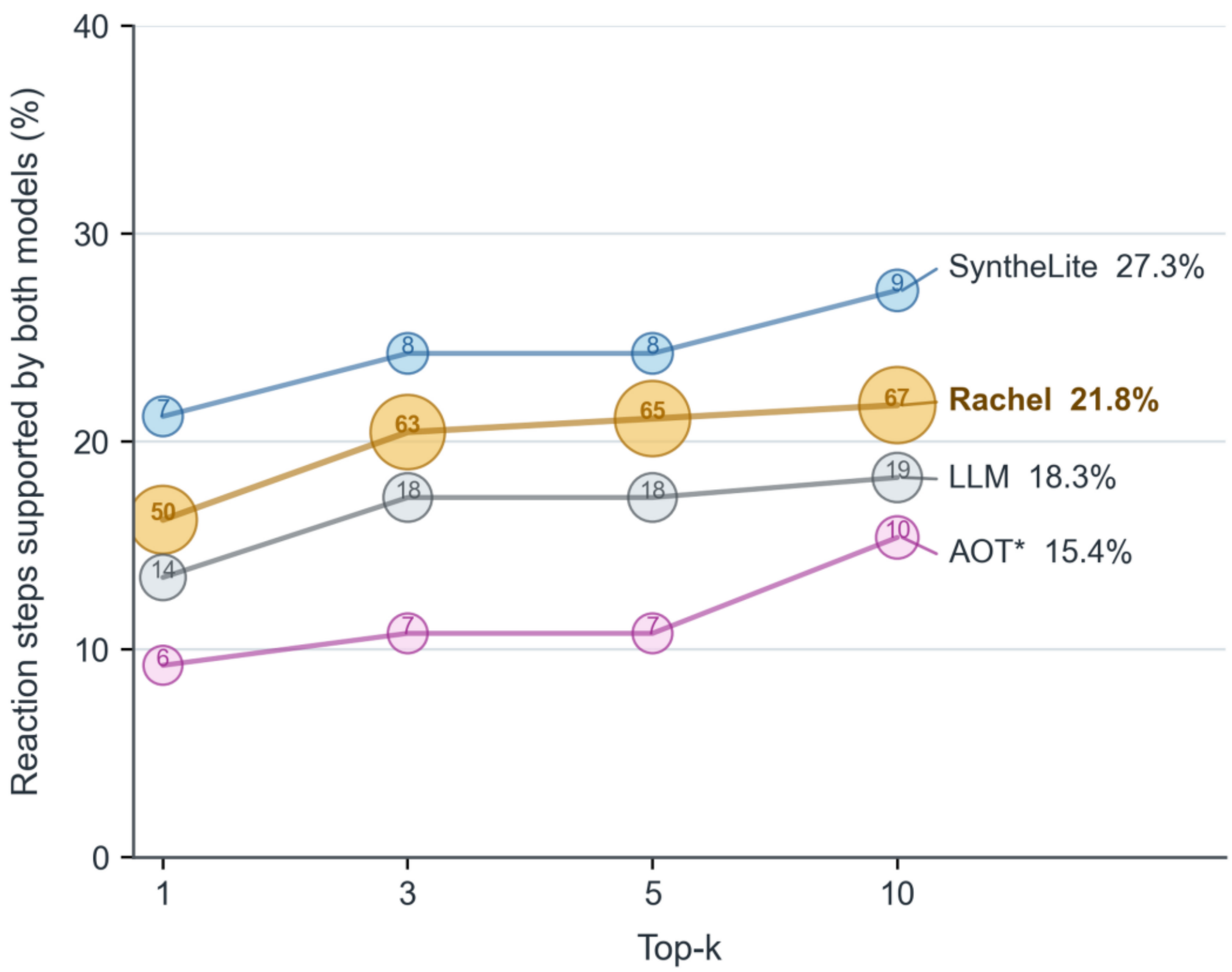

**Figure S19 | Forward-model agreement on method-specific RF25 routes.** Both@k requires ReactionT5v2-forward and RXNGraphormer each to recover the recorded product within their first k predictions (k = 1, 3, 5 and 10). Fractions pool equally weighted reaction steps within each method's strictly closed routes. Rachel, direct LLM (LLM), AOT* and SyntheLite contribute 24, 15, 14 and 6 routes, containing 308, 104, 65 and 33 jointly scored steps, respectively; all retained steps were scored by both models. Numerals indicate supported-step counts, with larger circles indicating more supported steps; end labels give Both@10 percentages. Lines connect k thresholds, not repeated runs. Route sets are unmatched across methods. Rachel has the largest absolute supported-step count, but the unmatched populations prevent interpreting this as a higher support fraction. This diagnostic reports predictive agreement under the specified inputs and matching rules. The RF25 values are lower than the common58 values and do not establish experimental correctness or feasibility (Supplementary Methods S4).